\documentclass[12pt,oneside,letterpaper]{book}

\usepackage{graphicx}

\begin{document}

\pagestyle{empty}

\begin{center}
{\LARGE APPLICATION OF HARMONIC MAPS \textit{CP}$^{(N-1)}$\\ON \textit{SU(N)} BOGOMOLNY EQUATION\\FOR BPS MAGNETIC MONOPOLES

\bigskip
} \vfill {\Large THESIS\\
As requirement for obtaining Master degree programme\\
from Bandung Institute of Technology}

\bigskip
%\vfill {\Huge\bf Sobre el momento magn'etico an'omalo del
%fot'on\\}
\end{center}\vfill\begin{flushleft}
{\Large \ \ \ \ \ \ \ \ \ \ \ \ \ \ \ \ Author:\ Ardian Nata Atmaja\\ \ \ \\
\bigskip
\begin{tabular}{@{}c@{}l@{}}
\ \ \ \ \ \ \ \ \ \ \ \ \ \ \ \  Supervisor:\ Hans Jacobus Wospakrik, Ph.D.\\
\end{tabular}
}
\bigskip
\bigskip
\bigskip
\bigskip
\bigskip
\bigskip
\vfill{
\begin{figure}[htbp]
	\centering
		\scalebox{2}{\includegraphics[width=2cm]{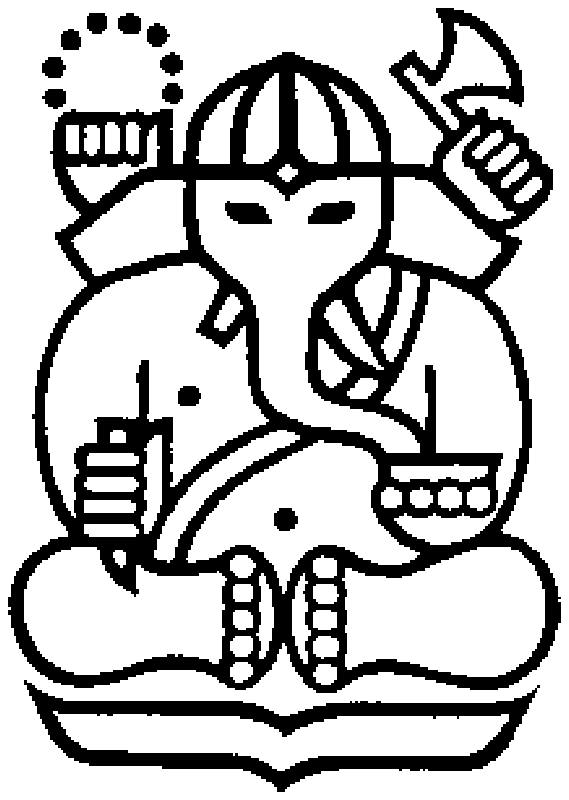}}
\end{figure}
}

\vfill
\begin{center}
{\Large
Department of Physics\\
Bandung Institute of Technology\\
Indonesia\\
2005\\
}\end{center}
\end{flushleft}
\clearpage

%*\begin{flushright}
%\textit{A mis padres}
%\end{flushright}

%\vfill \noindent``Lo esencial en la existencia de un hombre como
%yo es lo que 'el piensa y c'omo piensa; no lo que realice o
%sufra''
%\smallskip
%\begin{flushright}
%ALBERT EINSTEIN
%\end{flushright}
%\vfill
%\clearpage
\begin{flushright}

\textit{in memory of Hans Jacobus Wospakrik...}\\
\textit{(R.I.P. January 11th, 2005)}

\end{flushright}

\begin{figure}[htbp]
	\centering
	\includegraphics[width=5cm]{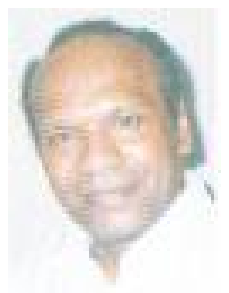}
\end{figure}

\clearpage

\pagestyle{plain}

\frontmatter

%****************************************** I N T R O D U C C I O N ***************************************************
\chapter{Abstract}

In this thesis we study dynamic of magnetic monopoles from Lagrangian
density in Yang-Mills-Higgs field theory. In particular, we discuss BPS
(Bogomolny Prasad Sommerfield) magnetic monopoles, described by \textit{SU(N)}
Bogomolny equations, which has field equations in form of non-linear coupled
matrix field equations. One of the methods to simplify \textit{SU(N)} Bogomolny
equations is by using harmonic maps $CP^{(N-1)}$. This method has relation with
\textit{Gr(n,N)} $\sigma$-model and can transform \textit{SU(N)} Bogomolny equation into more simple
scalar field equations that depends only on one variable. As an example, we consider the case of $SU(2)$ Bogomolny equation.

\tableofcontents

\mainmatter

\chapter{Introduction}

Magnetic monopoles concept was first introduced by P.A.M. Dirac while
he tried to search for explanation about the unit of electronic charge $e$~\cite{dirac}. His
research was based on the fact that electric charge is always observed in integral
multiples of the electronic charge $e$. This electronic charge $e$ has made $\frac{\hbar c}{e^{2}}$ approximately 137 which then became his focus of research and some others physicist, one of them was A.S.
Eddington. In his research development, Dirac considered some arguments but
unfortunately did not lead him to any value of 137 and for that reason he felt that
his arguments were a failure. Instead, the result of his research, which he wrote it
in a paper titled by \textit{Quantized Singularities in The Electromagnetic Field}, born the
new idea of magnetic monopoles. It was the concept that for years later interest
many scientists with capable of wide generalizations.

In the following years, many papers have been published in the topic of
magnetic monopoles. Between year 1973 through 1976, there were more than 300
research papers on the subject of magnetic monopoles~\cite{carrigan}. A primary contribution
for the theoretical investigations was provided by 't Hooft and Polyakov in 1974
that they discovered a magnetic monopoles solution in a spontaneously broken
non-Abelian gauge theory~\cite{hooft}. Their work pointed out the natural manner in which
magnetic monopoles make their appearance in these theories and encouraged
further exploration of this phenomenon.

Many studies about magnetic monopoles in non-Abelian gauge theory
were motivated by two sources. First, if magnetic monopoles solitons occurs in a
unified gauge theory of the weak and electromagnetic interactions, which then the
theory describes the real world proved to be correct, then they will be
experimentally accessible even though at extremely high energies. In this case,
there must be available information about these soliton of magnetic monopoles
field configurations. Second, there may be a connection between magnetic
monopoles and the quark confinement mechanism. These magnetic monopoles would probably be counter parts of some charge other than electric charge and therefore be expected to have little connection with the magnetic monopoles
sought in the experiments~\cite{marciano}.

In this thesis, we describe how to develop \textit{SU(N)} Bogomolny equations for
BPS magnetic monopoles from Yang-Mills-Higgs field theory and applying a
mathematical device, harmonic maps \textit{CP}$^{(N-1)}$, to transform the equations in the
more simple form that can be used for further methods to find the solutions. In the
first chapter, we discuss the Yang-Mills-Higgs field in Riemann sphere and derive
its dynamics equations. Then by using its energy equation, we take some
conditions and Bogomolny analysis to get \textit{SU(N)} Bogomolny equations for BPS
magnetic monopoles. For Chapter two, we investigate on harmonic maps and how
its connection with \textit{Gr(n,N)} $\sigma$-model. We take special case for \textit{CP}$^{(N-1)}$ space of
\textit{Gr(n,N)} $\sigma$-model and derive many of its properties and awe also use a Veronese
map as we will used for next chapter to obtain the solutions. Chapter three is the
major work of this thesis where we use ansatz for \textit{SU(N)} Bogomolny equations
in chapter one to connect with harmonic maps method in chapter two. With that
ansatz, we can transform the \textit{SU(N)} Bogomolny equations form its matrix form
into scalar equations which are much more simple to be discussed. We also take
an example for $SU(2)$ Bogomolny equations and find its solutions. The last
chapter is the conclusion about the results that we get in the previous chapters.

\chapter{Magnetic Monopoles}
\section{Dynamics of Yang-Mills-Higgs Field}
In searching for quantization of a unit of electric charge $e$, Dirac start his
attempt by writing Maxwell equations in matter (with electric charge and current
source) into symmetrical form between electric and magnetic field. This is based
of the fact that in vacuum conditions, Maxwell equations has symmetrical form
between its fields. So, why it does not occur in the same way as we move into
Maxwell equations in matter. For that reason, Dirac introduces a magnetic charge
and current source in his version of Maxwell equations~\cite{dirac}. Next, this idea was
extended in more general because as we know that Maxwell equations in vacuum
basically can be derived from pure $U(1)$ Yang-Mills-Higgs field theory.

In this section, we discuss about magnetic monopoles which is derived
from Yang-Mills-Higgs field theory with Lagrangian density given by~\cite{sutcliffe}
\begin{equation}
	\mathcal{L}=\frac{1}{8}tr\left(F_{\mu\nu}F^{\mu\nu}\right)-\frac{1}{4}tr\left(D_{\mu}\Phi D^{\mu}\Phi\right)-\frac{\lambda}{8}\left(\left\|\Phi\right\|^{2}-1\right)^{2}
\end{equation}
where $A_{\mu}$ is the gauge field and $\Phi$ is a Higgs field, as sources for magnetic monopoles, with $F_{\mu\nu}=\partial_{\mu}A_{\nu}-\partial_{\nu}A_{\mu}+\left[A_{\mu},A_{\nu}\right]$ and covariant derivative $D_{\mu}=\partial_{\mu}+\left[A_{\mu},\right]$. We also write the Higgs field in $\Phi=i\Phi^{a}T_{a}$ which is a skew Hermitian form so that $\left\|\Phi\right\|^{2}=-\frac{1}{2}tr\left(\Phi\Phi\right)$ and with $tr\left(T_{a}T_{b}\right)=2\delta_{ab}$ ($T_{a}$ is generator of Lie group $SU(N)$). For this chapter, we use
index convention $\mu,\nu=1,2,3$ and $a,b=1,\ldots,dim[SU(N)]$. If we look at the equation (2.1), we recognize than on the first part of equation (2.1) is for electric and magnetic field while the second part is for the sources of
magnetic monopoles and then the third part is a Lagrange multiplier that comes from constraint condition $\left\|\Phi\right\|^{2}=1$.
In order to write dynamics equations from Lagrangian density above, we have to take variation of action to each of its fields. As variation of action is wrote by
\begin{equation}
	\delta S=\int\delta \mathcal{L}\ d^{4}x
\end{equation}
If we look for dynamics equations of $A_{\mu}$ field then we take variation of $A_{\mu}$ to field to the equation (2.2) and action on Lagrangian becomes
\begin{eqnarray}
\delta \mathcal{L} & = &\frac{1}{4}tr(\partial_{\mu}\left(F^{\mu\nu}\delta A_{\nu}\right)-\partial_{\nu}\left(F^{\mu\nu}\delta A_{\mu}\right)-\partial_{\mu}F^{\mu\nu}\delta A_{\nu}+\partial_{\nu}F^{\mu\nu}\delta A_{\mu}-F^{\mu\nu}\left[A_{\nu},\delta A_{\mu}\right] \nonumber \\
 & & \mbox{} +F^{\mu\nu}\left[A_{\mu},\delta A_{\nu}\right])+\frac{1}{2}tr\left(D^{\mu}\Phi\left[\Phi,\delta A_{\mu}\right]\right)
\end{eqnarray}
Because we have boundary condition for integral on surface $\delta A_{\mu}=0$, then the second part of equation (2.3) can be ignored so that
\begin{equation}
	\delta \mathcal{L} =\frac{1}{2}tr((D_{\nu}F^{\mu\nu}+[D^{\mu}\Phi,\Phi])\delta A_{\mu})
\end{equation}
from least action principal $\delta S=0$, then dynamics equation for fields $A_{\mu}$ are
\begin{equation}
	D_{\nu}F^{\mu\nu}+[D^{\mu}\Phi,\Phi]=0
\end{equation}
While for $\Phi$ field, we take variation of $\Phi$ field to the equation (2.2) and the action on Lagrangian
becomes
\begin{eqnarray}
	\delta \mathcal{L}&=&-\frac{1}{2}tr(\partial_{\mu}(D^{\mu}\Phi\delta\Phi)-\partial_{\mu}D^{\mu}\Phi\delta\Phi+(D^{\mu}\Phi A_{\mu}\delta\Phi-D^{\mu}\Phi\delta\Phi A_{\mu})) \nonumber \\
	&&+\frac{\lambda}{4}(||\Phi||^{2}-1)tr(\Phi\delta\Phi)
\end{eqnarray}
and also for integral on surface we have boundary condition $\delta\Phi=0$, then the second part of equation (2.6) can be ignored so that
\begin{equation}
	\delta \mathcal{L}=\frac{1}{2}tr((D_{\mu}D^{\mu}\Phi+\frac{\lambda}{2}(||\Phi||^{2}-1)\Phi)\delta\Phi)
\end{equation}
From least action principal $\delta S=0$, then dynamics equation of $\Phi$ field is
\begin{equation}
	D_{\mu}D^{\mu}\Phi+\frac{\lambda}{2}(||\Phi||^{2}-1)\Phi)=0
\end{equation}
As we can see in equation (2.5) which is similar to Maxwell equations in matter with current sources is in form of field $\Phi$.

\section{Energy Equation for Yang-Mills-Higgs Field}
Now, let we find energy equation for Lagrangian (2.1) and then use it to get $SU(N)$ Bogomolny equations. To do that, we have to count the tensor energy-momentum for Yang-Mill-Higgs field by rewriting the action of Lagrangian (2.1) for arbitrary metric
\begin{equation}
	S=\int\sqrt{-g}\mathcal{L}\ d^{4}x
\end{equation}
where $g=det(g{\mu\nu})=e^{tr(ln\ g_{\mu\nu})}$. Then, we take variation over $g_{\mu\nu}$ to the action (2.9) so that
\begin{equation}
	\delta S=\int(\delta(\sqrt{-g})\mathcal{L}+\sqrt{-g}\delta \mathcal{L})d^{4}x
\end{equation}
with
\begin{eqnarray}
	\delta\sqrt{-g}=\frac{1}{2}(-g)^{-\frac{1}{2}}(-\delta g) \nonumber \\
	\delta g=e^{tr(ln\ g^{\mu\nu})}tr(\frac{1}{g_{\mu\nu}}\delta g_{\mu\nu})=g\ tr(g^{\mu\nu}\delta g_{\mu\nu})
\end{eqnarray}
If we write in its components $\delta g=g\ g^{\mu\nu}\delta g_{\mu\nu}$, so that
\begin{equation}
	\delta\sqrt{-g}=\frac{1}{2}(-g)^{\frac{1}{2}}g^{\mu\nu}\delta g_{\mu\nu}
\end{equation}
While
\begin{equation}
	\delta \mathcal{L}=\frac{1}{4}g^{\nu\sigma}tr(F_{\rho\sigma}F_{\mu}{\nu})\delta g^{\mu\rho}-\frac{1}{4}tr(D_{\rho}\Phi D_{\mu}\Phi)\delta g^{\mu\rho}
\end{equation}
and using $\delta g^{\mu\rho}=-g^{\mu\alpha}\delta g_{\alpha\beta}g^{\beta\rho}$, then
\begin{equation}
	\delta \mathcal{L}=(-\frac{1}{4}g^{\nu\sigma}tr(F_{\rho\sigma}F_{\mu\nu})+\frac{1}{4}tr(D_{\rho}\Phi D_{\mu}\Phi))g^{\mu\alpha}g^{\beta\rho}\delta g_{\alpha\beta}
\end{equation}
Substitute equations (2.12) and (2.14) into equation (2.10), then
\begin{equation}
	\delta S=-\frac{1}{2}\sqrt{-g}\int(-g^{\alpha\beta}\mathcal{L}+\frac{1}{2}(g^{\nu\sigma}tr(F_{\rho\sigma}F_{\mu\nu})-tr(D_{\rho}\Phi D_{\mu}\Phi))g^{\mu\alpha}g^{\beta\rho})\delta g_{\alpha\beta}d^{4}x
\end{equation}
From general relativity for action of matter is $\delta S=-\frac{1}{2}\sqrt{-g}\int T^{\alpha\beta}\delta g_{\alpha\beta}d^{4}x$ so we have tensor energy-momentum
\begin{equation}
	T^{\alpha\beta}=-g^{\alpha\beta}\mathcal{L}+\frac{1}{2}(g^{\nu\sigma}tr(F_{\rho\sigma}F_{\mu\nu})-tr(D_{\rho}\Phi D_{\mu}\Phi))g^{\mu\alpha}g^{\beta\rho}
\end{equation}
and take its energy part $T^{00}=-g^{00}\mathcal{L}+\frac{1}{2}(g^{\nu\sigma}tr(F_{\rho\sigma}F_{\mu\nu})-tr(D_{\rho}\Phi D_{\mu}\Phi))g^{\mu 0}g^{0\rho}$. From definition, we know that the energy is written by $E=\int T^{00}\sqrt{g}d^{3}x$ for arbitrary metric. Below, we write some metrics where $ds^{2}=g_{\mu\nu}dx^{\mu}dx^{\nu}$ :
\begin{enumerate}
	\item Standard Minkowskian metric
\begin{equation}
	ds^{2}=dt^{2}-dx^{2}-dy^{2}-dz^{2}	
\end{equation}
	\item Minkowskian metric with spherical coordinates
\begin{equation}
	ds^{2}=dt^{2}-dr^{2}-r^{2}d\theta^{2}-r^{2}\sin^{2}{\theta}d\phi^{2}	
\end{equation}
	\item Minkowskian metric with Riemann sphere coordinates
\begin{eqnarray}
	ds^{2}=dt^{2}-dr^{2}-\frac{2r^{2}}{(1+|\xi|^{2})^{2}}(d\xi d\bar{\xi}+d\bar{\xi}d\xi) \nonumber \\
	g_{\mu\nu}=\left(
\begin{array}{cccc}
	1 & 0 & 0 & 0 \\
	0 & -1 & 0 & 0 \\
	0 & 0 & 0 & -\frac{2r^{2}}{(1+|\xi|^{2})^{2}} \\
	0 & 0 & -\frac{2r^{2}}{(1+|\xi|^{2})^{2}} & 0 \\
\end{array}\right)
\end{eqnarray}
derivation of metric (2.19) is in Appendix A.
\end{enumerate}
So, we have the energy part $T^{00}$ with metric (2.19) is
\begin{equation}
	T^{00}=-\mathcal{L}+\frac{1}{2}g^{\nu\sigma}tr(F_{0\sigma}F_{0\nu})-\frac{1}{2}tr(D_{0}\Phi D_{0}\Phi)
\end{equation}

\subsection{Static field and BPS condition}
For this thesis, we will only look for solutions of static fields and in BPS limit. In that case, we have for static field $(\partial_{0}\Phi=0,A_{0}=0)$ and with BPS condition on $\lambda=0$, then equation (2.20) becomes
\begin{equation}
T^{00}=-\frac{1}{8}g^{\mu\rho}g^{\nu\sigma}tr(F_{\rho\sigma}F_{\mu\nu})+\frac{1}{2}g^{ij}tr(F_{0i}F_{0j})+\frac{1}{4}g^{ij}tr(D_{i}\Phi D_{j}\Phi)
\end{equation}
where we use index $i$ and $j=1,2,3=r,\xi\bar{\xi}$ and for index $\mu,\nu,\rho$ and $\sigma=0,1,2,3=t,r,\xi\bar{\xi}$. We can calculate equation (2.21) separately for each field. For the first one, we calculate the part that contains $\Phi$ field which is written as below
\begin{equation}
	\frac{1}{4}g^{ij}tr(D_{i}\Phi D_{j}\Phi)=-\frac{1}{4}tr\left(D_{r}\Phi D_{r}\Phi+\frac{(1+|\xi|^{2})^{2}}{r^{2}}D_{\xi}\Phi D_{\bar{\xi}}\Phi\right)
\end{equation}
While for the others part that bear $A_\mu$ field which are
\begin{equation}
	-\frac{1}{8}g^{\mu\rho}g^{\nu\sigma}tr(F_{\rho\sigma}F_{\mu\nu})+\frac{1}{2}g^{ij}tr(F_{0i}F_{0j})
\end{equation}
we can separate it again as below
\begin{itemize}
	\item $\mu=0\rightarrow\nu\neq 0,\rho=0,\sigma\neq 0$\\
	$\Rightarrow -\frac{1}{8}g^{ij}tr(F_{0j}F_{0i})$
	\item $\mu\neq 0,\nu=0\rightarrow\sigma=0,\rho\neq 0$\\
	$\Rightarrow -\frac{1}{8}g^{ij}tr(F_{0i}F_{0j})$
	\item $\mu\neq 0,\nu\neq 0\rightarrow\sigma\neq 0,\rho\neq 0$\\
	$\Rightarrow -\frac{1}{8}g^{ik}g^{jl}tr(F_{kl}F_{ij})$
\end{itemize}
Joining again all of its parts of the equation (2.21), then we have
\begin{equation}
	T^{00}=-\frac{1}{4}tr\left(D_r\Phi D_r\Phi+\frac{(1+|\xi|^2)^2}{r^2}D_\xi\Phi D_{\bar{\xi}}\Phi\right)+\frac{1}{4}g^{ij}tr(F_{0i}F_{0j})-\frac{1}{8}g^{ik}g^{jl}tr(F_{kl}F_{ij})
\end{equation}

\subsection{Magnetic monopoles conditions}
For magnetic monopoles it means that the sources only produces magnetic field or we may write $F_{0i}=0$ (if $F_{0i}\neq 0$ then it is known as dyons,it means the sources produces two fields which are magnetic and electric field). With this condition then equation (2.24) becomes
\begin{equation}
	T^{00}=-\frac{1}{4}tr\left(D_r\Phi D_r\Phi+\frac{(1+|\xi|^2)^2}{r^2}D_\xi\Phi D_{\bar{\xi}}\Phi+\frac{(1+|\xi|^2)^2}{r^2}F_{r\xi}F_{r\bar{\xi}}+\frac{(1+|\xi|^2)^4}{4r^4}F_{\xi\bar{\xi}}F_{\bar{\xi}\xi}\right)
\end{equation}
So, we have equation of energy for BPS magnetic monopoles as below
\begin{equation}
	E=-\frac{1}{4}\int tr\left(\frac{2r^2}{(1+|\xi|^2)^2}D_r\Phi D_r\Phi+2D_\xi\Phi D_{\bar{\xi}}\Phi+2F_{r\xi}F_{r\bar{\xi}}+\frac{(1+|\xi|^2)^2}{2r^2}F_{\xi\bar{\xi}}F_{\bar{\xi}\xi}\right)d^3x
\end{equation}

\section{Magnetic Monopoles Equations of Motion}
From the energy (2.26), we can derive equations of motion for each field by using variation of energy about each field and least action principal $\delta E=0$. So, for $\Phi$ field, the variation of the energy (2.26) is
\begin{eqnarray}
	\delta E&=&-\frac{1}{4}\int tr\Big(\partial_r\left(\frac{2r^2}{(1+|\xi|^2)^2}D_r\Phi\delta\Phi\right)-\partial_r\left(\frac{2r^2}{(1+|\xi|^2)^2}D_r\Phi\right)\delta\Phi \nonumber \\
	&& +\frac{2r^2}{(1+|\xi|^2)^2}[D_r\Phi,A_r]\delta\Phi+\partial_\xi(D_{\bar{\xi}}\Phi\delta\Phi)-\partial_\xi(D_{\bar{\xi}}\Phi)\delta\Phi+[D_{\bar{\xi}}\Phi,A_\xi]\delta\Phi \nonumber \\
	&& +\partial_{\bar{\xi}}(D_\xi\Phi\delta\Phi)-\partial_{\bar{\xi}}(D_\xi\Phi)\delta\Phi+[D_\xi\Phi,A_{\bar{\xi}}]\delta\Phi\Big)d^3x
\end{eqnarray}
Because we have boundary condition for surface integral $\delta\phi=0$, then
\begin{equation}
	\delta E=-\frac{1}{4}\int tr\left(-D_r\left(\frac{2r^2}{(1+|\xi|^2)^2}D_r\Phi\right)\delta\Phi-D_\xi(D_{\bar{\xi}}\Phi)\delta\Phi-D_{\bar{\xi}}(D_\xi\Phi)\delta\Phi\right)d^3x
\end{equation}
and equation of motion of $\Phi$ field is
\begin{eqnarray}
-D_r\left(\frac{2r^2}{(1+|\xi|^2)^2}D_r\Phi\right)-D_\xi(D_{\bar{\xi}}\Phi)-D_{\bar{\xi}}(D_\xi\Phi)=0 \\
D_r\left(r^2D_r\Phi\right)=-\frac{(1+|\xi|^2)^2}{2}\left(D_\xi(D_{\bar{\xi}}\Phi)+D_{\bar{\xi}}(D_\xi\Phi)\right)
\end{eqnarray}
With the same manner for $A_{\xi}$ field, then variation of the energy becomes
\begin{eqnarray}
\delta E&=&-\frac{1}{4}\int tr\Big(-[D_{\bar{\xi}}\Phi,\Phi]\delta A+\partial_r(F_{r\bar{\xi}}\delta A)-\partial_rF_{r\bar{\xi}}\delta A+[F_{r\bar{\xi}},A_r]\delta A \nonumber \\
&&-\partial_{\bar{\xi}}\left(\frac{(1+|\xi|^2)^2}{2r^2}F_{\bar{\xi}\xi}\delta A\right)+\partial_{\bar{\xi}}\left(\frac{(1+|\xi|^2)^2}{2r^2}F_{\bar{\xi}\xi}\right)\delta A \nonumber \\
&&-\left[\frac{(1+|\xi|^2)^2}{2r^2}F_{\bar{\xi}\xi},A_{\bar{\xi}}\right]\delta A\Big)d^3x
\end{eqnarray}
and using boundary condition for surface integral $\delta A=0$, then
\begin{equation}
\delta E=-\frac{1}{4}\int tr\left(-[D_{\bar{\xi}}\Phi,\Phi]\delta A-D_rF_{r\bar{\xi}}\delta A+D_{\bar{\xi}}(\frac{(1+|\xi|^2)^2}{2r^2}F_{\bar{\xi}\xi})\delta A\right)d^3x	
\end{equation}
So, equation of motion for $A_{\xi}$ field is
\begin{eqnarray}
-[D_{\bar{\xi}}\Phi,\Phi]-D_rF_{r\bar{\xi}}+D_{\bar{\xi}}\left(\frac{(1+|\xi|^2)^2}{2r^2}F_{\bar{\xi}\xi}\right)=0 \\ \left[D_{\bar{\xi}}\Phi,\Phi\right]+D_rF_{r\bar{\xi}}=\frac{1}{2r^2}D_{\bar{\xi}}\left((1+|\xi|^2)^2F_{\bar{\xi}\xi}\right)
\end{eqnarray}
If we do for $A_{\bar{\xi}}$ field then we have the same equation of motion above just by changing the index $\xi\leftrightarrow\bar{\xi}$ from equation (2.34), so the equation of motion for field $A_\xi$ is
\begin{equation}
\left[D_{\xi}\Phi,\Phi\right]+D_rF_{r\xi}=\frac{1}{2r^2}D_{\xi}\left((1+|\xi|^2)^2F_{\xi\bar{\xi}}\right)	
\end{equation}
While for $A_r$ field, if we take its variation on the energy, then
\begin{eqnarray}
\delta E&=&-\frac{1}{4}\int tr\Big(-\frac{2r^2}{(1+|\xi|^2)^2}[D_r\Phi,\Phi]\delta A+\partial_{\bar{\xi}}(F_{\xi r}\delta A)-\partial_{\bar{\xi}}F_{\xi r}\delta A+[F_{\xi r},A_{\bar{\xi}}]\delta A \nonumber \\
&& +\partial_\xi (F_{\bar{\xi}r}\delta A)-\partial_\xi F_{\bar{\xi}r}\delta A+[F_{\bar{\xi}r},A_\xi]\delta A\Big)d^3x
\end{eqnarray}
and with boundary condition for surface integral $\delta A_r$, then equation (2.36) becomes
\begin{equation}
\delta E=-\frac{1}{4}\int tr\left(-\frac{2r^2}{(1+|\xi|^2)^2}[D_r\Phi,\Phi]\delta A-D_{\bar{\xi}}F_{\xi r}\delta A-D_\xi F_{\bar{\xi}r}\delta A\right)d^3x	
\end{equation}
So, equation of motion for $A_r$ field is
\begin{eqnarray}
-\frac{2r^2}{(1+|\xi|^2)^2}[D_r\Phi,\Phi]-D_{\bar{\xi}}F_{\xi r}-D_\xi F_{\bar{\xi}r}=0 \\
\left[D_r\Phi,\Phi\right]=\frac{(1+|\xi|^2)^2}{2r^2}\left(D_{\bar{\xi}}F_{r\xi}+D_\xi F_{r\bar{\xi}}\right)	
\end{eqnarray}

\subsection{$SU(N)$ Bogomolny equation}
In the previous, we have derive equations of motion (2.30), (2.34), (2.35), and (2.39) which are known as $SU(N)$ non-Bogomolny equations for BPS magnetic monopoles. Those equations are in second order differential and have properties of non-linear and coupled fields, and it is hard to find solutions for that kind of equations. Next, we will derive $SU(N)$ Bogomolny equations for BPS magnetic monopoles which have equations in form of first order differential. Those kind of equations are more simple then $SU(N)$ non-Bogomolny equations and it is possible to find the solutions. In deriving $SU(N)$ Bogomolny equations, we use Bogomolny analysis for BPS magnetic monopoles to the energy equation (2.26). In Bogomolny analysis, then energy equation (2.26) is transformed into equation that contains square sum of the fields. We must consider that the energy has to have real value $E\geq0$, so that :
\begin{enumerate}
	\item For first part of the energy
\begin{eqnarray}
	-D_r\Phi &=&\left(D_r\Phi\right)^\dagger \nonumber \\
	&=&-\partial_r\Phi+[A_r^\dagger,\Phi]
\end{eqnarray}
then $A_r^\dagger=-A_r$.
	\item and second part
	\begin{eqnarray}
		-D_\xi\Phi &=&\left(D_{\bar{\xi}}\Phi\right)^\dagger \nonumber \\
		&=&-\left(\partial_\xi\Phi-[A_{\bar{\xi}}^\dagger,\Phi]\right)
	\end{eqnarray}
then $A_{\bar{\xi}}^\dagger=-A_\xi$.
	\item and third part is correct by using the result from 1) and 2)
	\begin{eqnarray}
		-F_{r\xi} &=&\left(F_{r\bar{\xi}}\right)^\dagger \nonumber \\
		&=&-\left(\partial_r A_\xi-\partial_\xi A_r+[A_r,A_\xi]\right)
	\end{eqnarray}
	\item also for fourth part is correct if we use the result from 1) and 2)
	\begin{eqnarray}
		-F_{\xi\bar{\xi}} &=&\left(F_{\bar{\xi}\xi}\right)^\dagger \nonumber \\
		&=&-\left(\partial_\xi A_{\bar{\xi}}-\partial_{\bar{\xi}}A_\xi+[A_\xi,A_{\bar{\xi}}]\right)
	\end{eqnarray}
\end{enumerate}
Let we look for some form of equations below:
\begin{equation}
(iD_\xi\Phi-F_{r\xi})(iD_\xi\Phi-F_{r\xi})^\dagger=-D_\xi\Phi D_{\bar{\xi}}\Phi-F_{r\xi}F_{r\bar{\xi}}+iD_\xi\Phi F_{r\bar{\xi}}-iF_{r\xi}D_{\bar{\xi}}\Phi
\end{equation}
\begin{eqnarray}
\left(iD_r\Phi-\frac{(1+|\xi|^2)^2}{2r^2}F_{\xi\bar{\xi}}\right)\left(iD_r\Phi-\frac{(1+|\xi|^2)^2}{2r^2}F_{\xi\bar{\xi}}\right)^\dagger&=&-D_r\Phi D_r\Phi-\frac{(1+|\xi|^2)^4}{4r^4}F_{\xi\bar{\xi}}F_{\bar{\xi}\xi} \nonumber \\
	&&+i\frac{(1+|\xi|^2)^2}{2r^2}D_r\Phi F_{\bar{\xi}\xi} \nonumber \\
	&&-i\frac{(1+|\xi|^2)^2}{2r^2}F_{\xi\bar{\xi}}D_r\Phi
\end{eqnarray}
then we can transform energy equation (2.26) so that it contains equations (2.44) and (2.45), and we may write it as
\begin{eqnarray}
	E&=&\frac{1}{4}\int tr\Big(\frac{(1+|\xi|^2)^2}{r^2}(iD_\xi\Phi-F_{r\xi})(iD_{\bar{\xi}}\Phi+F_{r\bar{\xi}})\nonumber \\
	&&+\left(iD_r\Phi-\frac{(1+|\xi|^2)^2}{2r^2}F_{\xi\bar{\xi}}\right)\left(iD_r\Phi+\frac{(1+|\xi|^2)^2}{2r^2}F_{\bar{\xi}\xi}\right) \nonumber \\
	&&+\left(i\frac{(1+|\xi|^2)^2}{r^2}F_{r\xi}D_{\bar{\xi}}\Phi\right)^\dagger+i\frac{(1+|\xi|^2)^2}{r^2}F_{r\xi}D_{\bar{\xi}}\Phi \nonumber \\
	&&+\left(i\frac{(1+|\xi|^2)^2}{2r^2}D_r\Phi F_{\bar{\xi}\xi}\right)^\dagger+i\frac{(1+|\xi|^2)^2}{2r^2}D_r\Phi F_{\bar{\xi}\xi}\Big)\frac{2r^2}{(1+|\xi|^2)^2}d^3x
\end{eqnarray}
This energy (2.46) has minimum a value if
\begin{eqnarray}
	iD_\xi\Phi=F_{r\xi} \\
	iD_r\Phi=\frac{(1+|\xi|^2)^2}{2r^2}F_{\xi\bar{\xi}}
\end{eqnarray}
and those equations are known as SU(N) Bogomolny equations for BPS magnetic
monopoles in Riemann sphere metric (2.19).

\chapter{Harmonic Maps}
\section{Harmonic Maps Definition}
Basically, harmonic map is a map between Riemannian manifolds. The
theory of harmonic maps was introduced in 1945 by B.F. Fuller and developed by
J. Eeels and J. M. Sampson ten years later~\cite{eels}. Its role in physics was shown
firstly by C. W. Misner~\cite{misner} in 1978, who formulated non-linear $\sigma$-model field
theory in geometrical description. The non-linear $\sigma$-models in two dimensions are
special interest because they bear many similarities to the non-abelian gauge
theory in four dimensions and have a property of being an integrable system.
Later in this thesis, the non-linear $\sigma$-model is shortly named $\sigma$-model.

Next, we explain some basic definitions of harmonic maps theory. Let $\mathcal{M}_0$ is a Riemannian manifold (source manifold) with local coordinates $u^i$ and metric
\begin{equation}
	ds^2=g_{ij}du^idu^j
\end{equation}
with $i,j=1,2,\ldots,dim[\mathcal{M}_0]$ and $\mathcal{M}$ is another Riemannian manifold (target manifold) with local coordinates $f^A$ with metric
\begin{equation}
	ds^2=h_{AB}df^Adf^B
\end{equation}
with $A,B=1,2,\ldots,dim[\mathcal{M}]$

Then, a map
\begin{eqnarray}
	f:\mathcal{M}_0\rightarrow\mathcal{M} \nonumber \\
	f=f(u)
\end{eqnarray}
is called a harmonic maps if it extremizes the action
\begin{equation}
	S=\int_{\mathcal{M}_0}d^{n_0}u\sqrt{g}\mathcal{L}
\end{equation}
such that $\delta_fS=0$ with $n=dim[\mathcal{M}_0]$ and $g=det(g_{ij})$ where
\begin{equation}
	\mathcal{L}=\frac{1}{2}h_{AB}\frac{\partial f^A}{\partial u^i}\frac{\partial f^B}{\partial u^j}g^{ij}
\end{equation}
is the Lagrangian density.

\section{$\sigma$-Model}
The theory of harmonic maps in physics literature is related to $\sigma$-model.
The $\sigma$-model is a field theory with the following properties~\cite{balachandaran}:
\begin{enumerate}
	\item Fields $\phi(u)$ of $\sigma$-model are constraints for all points
$\phi(u),\forall u\in\mathcal{M}_0$.
	\item The constrains and Lagrangian density are invariant under action of global
symmetry group $G$ on $\phi(u)$.
\end{enumerate}

\subsection{$O(N)\ \sigma$-model}
As an example, we look on $O(N)\ \sigma$-model which consist of $N$-real scalar
fields $\phi^{\hat{A}}$ with $\hat{A}=1,\ldots,N$ and has Lagrangian density
\begin{equation}
	\mathcal{L}=\frac{1}{2}\frac{\partial\phi^{\hat{A}}}{\partial u^i}\frac{\partial\phi^{\hat{A}}}{\partial u^j}g^{ij}
\end{equation}
where scalar fields $\phi^{\hat{A}}$ satisfy the constraints
\begin{equation}
\phi^{\hat{A}}\phi^{\hat{A}}=1
\end{equation}
with repeated index means the sum of all its value. Lagrangian density (3.6) is
invariant under global transformation group $O(N)$
\begin{equation}
\phi^{\hat{A}}\rightarrow(\phi^{\hat{A}})^\prime=O^{\hat{A}}_{\hat{B}}\phi^{\hat{B}}
\end{equation}
Geometrically, the constraints (3.7) define a sphere $(N-1)$ dimension $S^{N-1}$ in $N$ dimension Euclidean space $R^N$ of fields manifold $\phi^{\hat{A}}$. This
constraints can be solved by introducing parametrization below:
\begin{equation}
	\phi^A=f^A,\ \phi^N=\pm\sqrt{1-|f|^2},\ A=1,2,\ldots,(N-1)
\end{equation}
where $|f|^2=f^Af^A$ and the range of $|f|$ is limited to $-1\leq |f|\leq 1$. The choice of
sign + or – in equations (3.9) determines parametrization whether we use upper or
lower hemisphere of $S^{N-1}$. By parametrization in (3.9),
we can write Lagrangian density (3.6) as
\begin{equation}
	\mathcal{L}=\frac{1}{2}\frac{\partial\phi^{\hat{A}}}{\partial u^i}\frac{\partial\phi^{\hat{A}}}{\partial u^j}g^{ij}
\end{equation}
with $h_{AB}$ is tensor metric of target manifold $S^(N-1)$. The tensor metric $h_{AB}$ is defined
by substituting parametrization (3.9) into metric
\begin{eqnarray}
	d^2\sigma&=&d\phi_{\hat{A}}d\phi^{\hat{A}} \nonumber \\
	&=&\left(\delta_{AB}+\frac{f_Af_B}{1-|f|^2}\right)df^Adf^B
\end{eqnarray}
so that
\begin{equation}
	h_{AB}=\delta_{AB}+\frac{f_Af_B}{1-|f|^2}
\end{equation}
where $\delta_{AB}$ is the Kronecker delta. As the result, the
fields $f$ are free of constraints and are solution of
$O(N)\ \sigma$-model field, so its defines a harmonic maps
\begin{equation}
	f:\mathcal{M}_0\rightarrow S^{N-1}.
\end{equation}

\subsection{Group formulation in $\sigma$-model}
In group formulation, target manifold $\mathcal{M}$ of a
$\sigma$-model, with $G$ is the invariant global group, are homogeneous space of $G$. It
means that by applying transformation group $G$ over a field $\phi_p\in\mathcal{M}$ then we can get into all the fields in target manifold $\mathcal{M}$. In another words, for an arbitrary field $\phi_q\neq\phi_p\in\mathcal{M}$ at least there is an element $g\in G$ such that $\phi_q=g\phi_p$.

If there is any stability or isotropy group $H\subset G$
on the fields $\phi_p\in\mathcal{M}$
\begin{equation}
	H=\{ h\in G| h\phi_p=\phi_p \}
\end{equation}
then target Manifold $\mathcal{M}$ is a coset space of $G/H=\{gh|g\in G\}$ that works on the field $\phi_p$. If the
identity $I_G$ is the only subgroup of $G$, then $\mathcal{M}=G/H=G$ is manifold of group $G$, then we call it as chiral model.

In general, the target manifold $\mathcal{M}$ of $\sigma$-model
is a manifold of group $G$. In that case, the $\sigma$-model can be represented in group $G$
by writing parameters of group $G$ as the fields $f^A,\ A=1,\ldots,\left(n=dim[G]\right)$,
with $dim[G]$ is dimension of group $G$ or number of generator in group $G$. Hence,
the tensor metric for target manifold in $\sigma$-model is written by
\begin{equation}
	h_{AB}=-2tr\left(G^{-1}\frac{\partial G}{\partial f^A}G^{-1}\frac{\partial G}{\partial f^B}\right)
\end{equation}
and the Lagrangian density is
\begin{equation}
	\mathcal{L}=-tr\left(G^{-1}\frac{\partial G}{\partial f^A}G^{-1}\frac{\partial G}{\partial f^B}\right)
\end{equation}

\section{Grassmanian $\sigma$-Model}
Grassmannian manifold $Gr(n,N),\ 1\leq n<N$, is the
manifold of $n$ dimensional planes passing through the origin in the $N$ dimensional
complex space $C^N$, where $Gr(1,N)=CP^{N-1}$. So, the $CP^{N-1}$ is a set of lines
passing through the origin in $N$ dimensional complex space $C^N$ such that a point in $CP^{N-1}$ is a line
on $C^N$. In this thesis, we use Grassmannian manifold in complex coordinates.

In coset space formulation, the Grassmannian manifold is
\begin{equation}
	Gr(n,N)=\frac{U(N)}{U(N-n)\times U(n)}
\end{equation}
and if we take into account the orientation of the planes, then
\begin{equation}
	Gr(n,N)=\frac{SU(N)}{SU(N-n)\times SU(n)}
\end{equation}
$Gr(n,N)\ \sigma$-Model consists of $(N\times n)$ complex matrix fields $Z=(Z^{\hat{A}a})$ with $\hat{A}=1,\ldots,N$
and $a=1,\ldots,n$. Those fields satisfy the constraint:
\begin{equation}
	Z^\dagger Z=I_n
\end{equation}
The Lagrangian density of the model is required to be invariant under global
unitary transformations $G\in U(N)$ that work on $Z$ field from the left
\begin{equation}
	Z\rightarrow Z^\prime=GZ
\end{equation}
and also invariant under local gauge transformations $H(x)\in U(n)$ from the right of $Z$ field
\begin{equation}
	Z\rightarrow Z^\prime=ZH(x)
\end{equation}
In this thesis, we use Lagrangian density given by W. J. Zakrzewski~\cite{zakrzewski}
\begin{equation}
	\mathcal{L}=tr\left((D^\mu Z)^\dagger D_\mu Z\right)
\end{equation}
with
\begin{equation}
	D_\mu Z=\partial_\mu Z-ZZ^\dagger\partial_\mu Z
\end{equation}
Next, we define $w^a,\ a=1,\ldots,n$ to be a set of $N$ components orthonormal vectors in $C^N$, then the
matrix field $Z$ can be represented as
\begin{equation}
	Z=(w^1,\ldots,w^n)
\end{equation}
In this representation, $Z$ defines an orthonormal $n$-frame in $C^N$.

Let we choose $Z$ as the last $n$-columns of matrix $G$
\begin{equation}
	G=\left(Y^{\hat{A}B}\ Z^{\hat{A}a}\right),\ B=1,\ldots,(N-n)
\end{equation}
where $Y$ is an $\left(N\times (N-n)\right)$ matrix and $Z$ is an $(N\times n)$ matrix. Since we have unitary condition for $G\in U(N)$ which is $G^\dagger G=GG^\dagger=I_N$, then we have
\begin{equation}
	Y^\dagger Y=I_{N-n},\ Y^\dagger Z=0,\ Z^\dagger Z=I_n,\ YY^\dagger + ZZ^\dagger=I_N
\end{equation}
that makes $Z$ satisfies the constraint (3.19). For our purpose, we define field
\begin{eqnarray}
	\Phi&=&(Y\ Z)\left(\begin{array}{c}
	Y^\dagger \\
	Z^\dagger \\
	\end{array}\right)\nonumber \\
	&=&\{(Y\ Z)-(0\ 2Z)\}\left(\begin{array}{c}
	Y^\dagger \\
	Z^\dagger \\
	\end{array}\right)
\end{eqnarray}
and by using the property in equation (3.24), then we can write as
\begin{equation}
	\Phi=(I-2P)
\end{equation}
where $P=ZZ^\dagger$. A complete discussion of $Gr(n,N)\ \sigma$-model and for interested reader can study the reference on dissertation of Hans Jacobus Wospakrik~\cite{wospakrik}.

\section{$Gr(n, N)\ \sigma$-Model in Projection Space}
The Lagrangian density for the $Gr(n,N)\ \sigma$-model is simply given by W. J.
Zakrzewski~\cite{zakrzewski}
\begin{equation}
	\mathcal{L}=\frac{1}{8}tr(\partial_\mu\Phi\partial^\mu\Phi)
\end{equation}
with $\Phi=I-2P$ and $P$ is the projection operator, then
\begin{equation}
	\mathcal{L}=\frac{1}{2}tr(\partial_\mu P\partial^\mu\ P)
\end{equation}
Taking into account the constraint below:
\begin{eqnarray}
	P^2=P=P^\dagger \nonumber \\
	(P^2-P)=0
\end{eqnarray}
then we can write Lagrangian density as
\begin{equation}
	\mathcal{L}=\frac{1}{2}tr(\partial_\mu P\partial^\mu\ P)+\lambda tr(P^2-P)
\end{equation}
So, action on Lagrangian density (3.32) is
\begin{equation}
	S=\int\left(\frac{1}{2}tr(\partial_\mu P\partial^\mu\ P)+\lambda tr(P^2-P)\right)d^4x
\end{equation}

\subsection{Equations of Motion for $Gr(n,N)\ \sigma$-model}
Equations of motion of Lagrangian density (3.32) is searched by variating
the action (3.33) about $P$ and $\lambda$, then we get
\begin{equation}
	\delta S=\int tr\left(\partial^\mu(\partial_\mu P\delta P)-\partial^\mu\partial_\mu P\delta P+\lambda(2P-I)\delta P+(P^2-P)\delta\lambda\right)d^4x
\end{equation}
Because of boundary condition $\delta P=0$ on surface integral, we can discard the first part of equation (3.34). And then, we get the equations of motion as below
\begin{eqnarray}
	\partial^\mu\partial_\mu P-\lambda(2P-I)=0 \\
	P^2-P=0
\end{eqnarray}
We can see that the equation (3.36) returns to the constraint (3.31). Next, if we
multiply the equation (3.35) by $P$ on the right and
left separately, then we have two equations
\begin{eqnarray}
	\partial^\mu\partial_\mu PP-\lambda(2P-I)P=0 \\
	P\partial^\mu\partial_\mu P-\lambda P(2P-I)=0
\end{eqnarray}
If we count the difference between equation (3.37) and equation (3.38) then use
property (3.31), then we get
\begin{equation}
	\left[P,\partial^\mu\partial_\mu P\right]=0
\end{equation}

\subsection{$Gr(n, N)\ \sigma$-model solution}
In this section, we discuss the construction of solutions of $Gr(n,N)\ \sigma$-model
from equation (3.39) in two dimensional Euclidean space $R^2$ or the complex plane $C$. Later,
we compactify $R^2$ by including points at $\infty$ to obtain the Riemann sphere $S^2=R^2\cup{\infty}$
and consider the harmonic maps: $S^2\rightarrow Gr(n,N)$. For next
discussion, we use a $Gr(n,N)\ \sigma$-model that has source manifold in two
dimensional Euclidean space.

In searching for the solutions, we use the complex coordinates $(\xi,\bar{\xi})$ for $Gr(n,N)\ \sigma$-model therefore the equation (3.39) becomes
\begin{equation}
	\left[P,\partial_\xi\partial_{\bar{\xi}}P\right]=0
\end{equation}
Derivation of equation (3.40) is in Appendix B. We also can write equation (3.40)
in form of
\begin{equation}
	\partial_\xi\left[P,\partial_{\bar{\xi}}P\right]+\partial_{\bar{\xi}}\left[P,\partial_\xi P\right]=0
\end{equation}
Complete derivation of this equation in in Appendix B.

\subsection{Instanton solution}
Next, we look at differentiation of operator property $P^2=P$ on $\partial_\xi(P^2=P)$ and $\partial_{\bar{\xi}}(P^2=P)$ then we get
\begin{eqnarray}
	\partial_\xi PP=\partial_\xi P-P\partial_\xi P \\
	P\partial_{\bar{\xi}}P=\partial_{\bar{\xi}}P-\partial_{\bar{\xi}}PP
\end{eqnarray}
while the equation (3.41) is written by
\begin{equation}
	\partial_\xi\left(P\partial_{\bar{\xi}}P-\partial_{\bar{\xi}}PP\right)+\partial_{\bar{\xi}}\left(P\partial_\xi P-\partial_\xi PP\right)=0
\end{equation}
Substitute equations(3.42) and (3.43) to equation (3.44), then we get
\begin{equation}
	\partial_{\bar{\xi}}\left(P\partial_\xi P\right)-\partial_\xi\left(\partial_{\bar{\xi}}PP\right)=0
\end{equation}
Special class of solutions for equation (3.45) is satisfied when $P\partial_\xi P=0$ such that $\left(P\partial_\xi P\right)^\dagger=\partial_{\bar{\xi}}PP=0$, so $P\partial_\xi P=0$ is the solution of field equation which is called \textit{selfdual} equation. If we do the same way by substituting $P\partial_\xi P$
and $\partial_{\bar{\xi}}PP$ in equations (3.42) and (3.423) into equation
(3.44), then it will produce a solution $P\partial_{\bar{\xi}}P=0$ which is
called \textit{anti-selfdual} equation.

In this thesis, we focus on solution from \textit{selfdual} equation $P\partial_\xi P=0$. For this purpose, we consider an un-normalized $(N\times n)$ matrix field $M=M(\xi,\bar{\xi})$ for which the $|M|^2=M^\dagger M$ is assumed to
be non-singular. As $|M|^2$ is Hermitian, its eigen values
are real and there exist an unitary matrix $U$ such that
\begin{equation}
	|M|^2=U^\dagger \Lambda^2 U
\end{equation}
where $\Lambda^2$ is a diagonal matrix with eigen values $(\lambda_1^2,\ldots,\lambda_N^2)$. We define $\Lambda$ to be the square root matrix of $\Lambda^2$, hence $\Lambda=\left(\Lambda^2\right)^\frac{1}{2}=(\lambda_1,\ldots,\lambda_N)$. So, matrix $|M|^2=U^\dagger\Lambda^2U$ can be denoted by $|M|^2=\left(U^\dagger\Lambda U\right)\left(U^\dagger\Lambda U\right)$ such that $|M|=\left(|M|^2\right)^\frac{1}{2}=U^\dagger\Lambda U$.

In terms of $M$, the matrix field $Z$ of $Gr(n,N)\ \sigma$-model is given by
\begin{equation}
	Z=M|M|^{-1}
\end{equation}
and so the $P$ projection operator is
\begin{equation}
	P=ZZ^\dagger=\left(M|M|^{-1}\right)\left(|M|^{-1}M^\dagger\right)=M|M|^2M^\dagger
\end{equation}
where we have use property of matrix $\left(|M|^{-1}\right)^\dagger=|M|^{-1}$

Using property of
\begin{equation}
	(I-P)M=0
\end{equation}
then the \textit{selfdual} equation is given by
\begin{equation}
	M|M|^{-2}\left(\partial_{\bar{\xi}}M\right)^\dagger(I-P)=0
\end{equation}
Complete derivation of equations (3.49) and (3.50) is available in Appendix C.

The solution of equation (3.50) is $\partial_{\bar{\xi}}M=0$ or $M=M_0(\bar{\xi})$ which is called a \textit{holomorphic} matrix field. This solution of \textit{selfdual} equation is named instanton solution. With the same analogy, if we discuss about \textit{anti-selfdual} equation then we will get \textit{anti-holomorphic}
matrix field $M=\bar{M}_0(\bar{\xi})$ which is named anti-instanton solution.

\section{Full Solutions of $Gr(n,N)\ \sigma$-Model}
In this section, we will discuss the method of generating more general
exact solutions of the two dimensional $Gr(n,N)\ \sigma$-model starting from an
instanton solution. The method was originally introduced by A. Din and W. J
Zakrzewski~\cite{din,zakrzewski}.

Let $M_k=M_k(\xi,\bar{\xi}),\ k=1,\ldots,\lambda$ where $\lambda\leq(N-1)$,be a set of $(\lambda+1)$ mutually orthogonal $(N\times n)$ matrices, with $n<N$
\begin{equation}
	M^\dagger_kM_l=|M_k|^2\delta_{kl}
\end{equation}
where
\begin{equation}
	|M_k|^2=M_k^\dagger M_k
\end{equation}
are $(n\times n)$ non-singular matrices. Then the corresponding
projection operator for each matrix $M_k$ is given by
\begin{equation}
	P_k(n)=M_k|M_k|^{-2}M_k^\dagger
\end{equation}
Clearly, $tr(P_k(n))=tr(I_n)$, which means that each projection
operator has rank-n. From equation (3.53), we can show that the projection
operators are mutually orthogonal and also Hermitian
\begin{eqnarray}
	P_k(n)P_l(n)=\delta_{kl}P_l(n) \\
	P_k(n)^\dagger=P_k(n)
\end{eqnarray}
where $|M|^2$ are Hermitian.
In the following we want to present a generalized harmonic maps
ansatz.To do this we use a sequence of mutually orthogonal matrices $(M_0,M_1,\ldots,M_\lambda)$ obtained from a sequence of holomorphic (analytic) matrices $(M,\partial_\xi M,\ldots,\partial_\xi^\lambda M),\ \partial_{\bar{\xi}}M=0$ through the process of Gram-Schmidt orthogonalization.

We can do this using the operator $P_+$ which is defined by its action on any matrix $M\in C^{N\times n}$~\cite{din,zakrzewski}
\begin{equation}
	P_+M=\partial_\xi M-M|M|^2M^\dagger\partial_\xi M
\end{equation}
Then we have $M_0=M,M_1=P_+M,\ldots,M_k=P_+^kM,\ldots,M_\lambda=P_+^\lambda M$ or simply
\begin{equation}
	M_0=M,\ M_k=(I-P_{k-1})\partial_\xi M_{k-1},\ k=1,\ldots,\lambda
\end{equation}
where $P_{k-1}$ is the projection operator (3.53). An equivalent formulation for the sequence $M_k$, in terms
of the projection operators $P_k$, is given by
\begin{equation}
	M_k=(1-P_0-\cdots-P_{k-1})\partial_\xi^kM_0
\end{equation}
With either one of these constructions in equations (3.57) and (3.58) the
following properties of the matrices $M_k$ hold when $M_0$ is holomorphic
\begin{eqnarray}
	\partial_{\bar{\xi}}M_k=-M_{k-1}|M_{k-1}|^{-2}|M_k|^2 \\
	\partial\left(M_k|M_k|^{-2}\right)=M_{k+1}|M_k|^{-2}
\end{eqnarray}
Detail derivation is in Appendix D.

Notice also that, for the $CP^{(N-1)}$ case, projection operators $P_k$ with $k=0,\ldots,(N-1)$ are complete
\begin{equation}
	P_0+P_1+\cdots+P_{N-1}=I
\end{equation}
and according to the construction $M_k$ in equation
(3.58), then
\begin{equation}
	M_N=0
\end{equation}

With the projection operators $P_k$ that we have constructed above, we have the following result that was originally proved by A. Din and W.J. Zakrzewski ~\cite{din} using the $Z$ fields formalism.

\textbf{Theorem}: Each $(N\times N)$ projection operator $P_k(n)=M_k|M_k|^{-2}M_k^\dagger$, $k=0,1,\ldots,\lambda$, where $M_k=P_+^kM_0$ and $M_0=M_0(\xi)$ is a holomorphic $(N\times n)$ matrix field, solves the $Gr(n,N)\ \sigma$-model (2.19) for its equation of motion (3.40).

As a proof, we use the method that was developed by Sasaki~\cite{sasaki}, where
we shall be using the projection operator in term of $M_k$ matrix field as follows:

From properties (3.58) and (3.59), we derive:
\begin{eqnarray}
	\partial_\xi P_0&=&\partial_\xi (M_0|M_0|^{-2})M_0^\dagger+M_0|M_0|^{-2}\partial_\xi M_0^\dagger \nonumber \\
	&=&M_1|M_0|^{-2}M_0^\dagger \\
	\partial_\xi P_k&=&\partial_\xi (M_k|M_k|^{-2})M_k^\dagger+M_k|M_k|^{-2}(\partial_{\bar{\xi}} M_k)^\dagger \nonumber \\
	&=&M_{k+1}|M_k|^{-2}M_k^\dagger-M_k|M_{k-1}|^{-2}M_{k-1}^\dagger
\end{eqnarray}
with $k=1,\ldots,\lambda$.

Define
\begin{equation}
	\mathcal{Q}_k=\sum_{l=0}^{k-1}P_l,\ \mathcal{Q}_0=0
\end{equation}
and take its differential about $\xi$ then use equations (3.63) and (3.64)
\begin{eqnarray}
	\partial_\xi\mathcal{Q}_k&=&\sum_{l=0}^{k-1}\partial_\xi P_l \nonumber \\
	&=&M_k|M_{k-1}|^{-2}M_{k-1}^\dagger
\end{eqnarray}
Using orthogonality property of $M_k$ we obtain
\begin{equation}
	\mathcal{Q}_k\partial_\xi\mathcal{Q}_k=0
\end{equation}
where $P_k$ satisfy
\begin{equation}
	\mathcal{Q}_k\partial_\xi P_k=0,\ P_k\partial_\xi\mathcal{Q}_k=\partial_\xi\mathcal{Q}_k
\end{equation}
Then define operator
\begin{equation}
	R_k=\mathcal{Q}_k+P_k
\end{equation}
which satisfy \textit{self-dual} equation
\begin{eqnarray}
	R_k\partial_xi K&=&\partial_\xi\mathcal{Q}_k+P_k\partial_\xi P_k \nonumber \\
	&=&0
\end{eqnarray}
and take the Hermitian conjugate of (3.70) gives
\begin{equation}
	\partial_{\bar{\xi}}\mathcal{Q}_k+\partial_{\bar{\xi}}P_kP_k=0
\end{equation}

If we take differential of equation (3.70) about $\xi$ and of equation (3.71) about $\bar{\xi}$, then we get
\begin{equation}
	\partial_{\bar{\xi}}\partial_\xi\mathcal{Q}_k+\partial_{\bar{\xi}}P_k\partial_\xi P_k+P_k\partial_{\bar{\xi}}\partial_\xi P_k=0
\end{equation}
and
\begin{equation}
	\partial_\xi\partial_{\bar{\xi}}\mathcal{Q}_k+\partial_{\bar{\xi}}P_k\partial_\xi P_k+\partial_{\bar{\xi}}\partial_\xi P_kP_k=0
\end{equation}
The difference between equation (3.72) and (3.73) gives us
\begin{equation}
	\left[P_k,\partial_\xi\partial_{\bar{\xi}}P_k\right]=0
\end{equation}
which completes the proof.

\section{Action of Full Solutions and Non-Abelian Toda Equations}
As each projection operator $P_k(n)$ solves the $Gr(n, N)$ $\sigma$-model as stated in previous theorem (3.74), each projection operator $P_k(n)$ describes a specific field configuration having the action
\begin{equation}
	S_k=i\int d\xi d\bar{\xi}\ tr\left(\partial_{\bar{\xi}}P_k\partial_\xi P_k\right)
\end{equation}
where $S_0$ corresponds to instanton(or anti-instanton) configuration.

Using relation relation in equations (3.59) and (3.60), we have
\begin{eqnarray}
	\partial_{\bar{\xi}}P_k&=&\partial_{\bar{\xi}}M_k|M_k|^{-2}M_k^\dagger +M_k\partial_{\bar{\xi}}\left(|M_k|^{-2}M_k^\dagger\right) \nonumber \\
	&=&-M_{k-1}|M_{k-1}|^{-2}M_k^\dagger+M_k|M_k|^{-2}M_{k+1}^\dagger
\end{eqnarray}
If we substitute equations (3.64) and (3.76) into equation (3.75), then the action
becomes
\begin{eqnarray}
	S_k&=&i\int d\xi d\bar{\xi}\ tr\left(|M_k|^2|M_{k-1}|^{-2}+|M_{k+1}|^2|M_k|^{-2}\right) \nonumber \\
	&=&2\pi\left(\mathcal{N}_k+\mathcal{N}_{k-1}\right)
\end{eqnarray}
where
\begin{equation}
	\mathcal{N}_k=\frac{i}{2\pi}\int d\xi d\bar{\xi}\ tr\left(|M_{k+1}|^2|M_k|^{-2}\right)
\end{equation}
and by definition $\mathcal{N}_{-1}=0$ because $\partial_{\bar{\xi}}M_0=0$.

In the following, we derive recurrence relations for $tr\left(|M_k|^2|M_{k-1}|^{-2}\right)$ appearing in the integral (3.78). To do this we rewrite the definition of $M_{k+1}$ in (3.57) as below:
\begin{equation}
	\partial_\xi M_k=M_{k+1}+P_k\partial_\xi M_k
\end{equation}
By using $\partial_{\bar{\xi}}M_k$ in equation (3.59) and property
\begin{eqnarray}
	M_k^\dagger\partial_{\bar{\xi}}M_k&=&-M_k^\dagger M_{k-1}|M_{k-1}|^{-2}|M_k|^2 \nonumber \\
	&=&0
\end{eqnarray}
then we take differential of (3.79) about $\bar{\xi}$ and multiply on the left by $M_k^\dagger$ and on the right by $|M_k|^{-2}$, and becomes
\begin{equation}
M_k^\dagger\partial_\xi\partial_{\bar{\xi}}M_k|M_k|^{-2}=M_k^\dagger\partial_{\bar{\xi}}M_{k+1}|M_k|^{-2}+M_k^\dagger\partial_{\bar{\xi}}\left(P_k\partial_\xi M_k\right)|M_k|^{-2}
\end{equation}
The left side of equation (3.81) gives
\begin{equation}
	M_k^\dagger\partial_\xi\partial_{\bar{\xi}}M_k|M_k|^{-2}=-|M_k|^2|M_{k-1}|^{-2}
\end{equation}
.From equation (3.59), the first part of the right side gives
\begin{eqnarray}
	M_k^\dagger\partial_{\bar{\xi}}M_{k+1}|M_k|^{-2}&=&-M_k^\dagger M_k|M_k|^{-2}|M_{k+1}|^2|M_k|^{-2} \nonumber \\
	&=&-|M_{k+1}|^2|M_k|^{-2}
\end{eqnarray}
and the second part becomes
\begin{equation}
	M_k^\dagger\partial_{\bar{\xi}}\left(P_k\partial_\xi M_k\right)|M_k|^{-2}=\partial_\xi\left(\partial_{\bar{\xi}}|M_k|^2|M_k|^{-2}\right)
\end{equation}
The detail of equations (3.82) and (3.84) is in Appendix E.

Substitute equations (3.82), (3,83), and (3.84) into equation (3.81) that gives
\begin{equation}
	\partial_\xi\left(\partial_{\bar{\xi}}|M_k|^2|M_k|^{-2}\right)=|M_{k+1}|^2|M_k|^{-2}-|M_k|^2|M_{k-1}|^{-2}
\end{equation}
For the $n=1$ or CP(N-1) case, when $M_k$ are s sequence of $N$-component
vector fields, the equation (3.85) gives the Toda equation~\cite{bolton1}. If we take the trace
of (3.85) then we obtain
\begin{equation}
tr\left(\partial_\xi\left(\partial_{\bar{\xi}}|M_k|^2|M_k|^{-2}\right)\right)=\partial_\xi\partial_{\bar{\xi}}tr\left(ln|M_k|^2\right)
\end{equation}
Because $det|M_k|^2=e^{tr\left(ln|M_k|^2|\right)}$, we may write equation (3.86) as
\begin{equation}
tr\left(\partial_\xi\left(\partial_{\bar{\xi}}|M_k|^2|M_k|^{-2}\right)\right)=\partial_\xi\partial_{\bar{\xi}}ln\left(det|M_k|^2\right)
\end{equation}
So, the trace of equation (3.85) is
\begin{equation}
\partial_\xi\partial_{\bar{\xi}}ln\left(det|M_k|^2\right)=tr\left(|M_{k+1}|^2|M_k|^{-2}\right)-tr\left(|M_k|^2|M_{k-1}|^{-2}\right)
\end{equation}
Note that, for $Gr(1,N)=CP^{N-1}$ case, the equation (3.79) becomes
\begin{equation}
\partial_\xi\partial_{\bar{\xi}}ln\left(|M_k|^2\right)=|M_{k+1}|^2|M_k|^{-2}-|M_k|^2|M_{k-1}|^{-2}
\end{equation}

\section{Veronese Map}
In this subsection, we describe how to construct the full solutions of the
two dimensional Grassmanian $\sigma$-model using Veronese map. Here, we consider
only for $Gr(1,N)=CP^{N-1}$ case with $M_0\in C^{N\times 1}$ is a N-component vector field in complex space
\begin{eqnarray}
	M_0:C\rightarrow C^N \nonumber \\
	\xi\rightarrow M_0=(f_0,\ldots,f_p,\ldots,f_{N-1})^T
\end{eqnarray}
In the Veronese map~\cite{bolton2}, we choose $f_p$ as function of $\xi$ in order $p$ such that
\begin{equation}
	|M_0|^2=\left(1+|\xi|^2\right)^{N-1}
\end{equation}
This condition restricts the component of $f_p$ to have the form $f_p=\sqrt{C_p^{N-1}}\xi^p$ where $C_p^{N-1}$ is the combinatorial factor and so the corresponding $CP^{N-1}$ field is
\begin{equation}
Z_0=\frac{M_0}{|M_0|}=\frac{\left(1,\ldots,\sqrt{C_p^{N-1}}\xi^p,\ldots,\xi^{N-1}\right)^T}{\sqrt{\left(1+|\xi|^2\right)^{N-1}}}
\end{equation}
For general case $N$ in the sequence of $M_k$, we use proof that was first given by Ioannidou et. al~\cite{ioannidou1}:

\textbf{Proposition}:For the Veronese map $M_0$ in (3.92), the sequence $M_k,\ (k=0,\ldots,N-1)$, constructed by equations (3.57) or (3.58), called Veronese sequence~\cite{bolton2}, satisfy
\begin{equation}
	\frac{|M_{k+1}|^2}{|M_k|^2}=\frac{(k+1)(N-k-1)}{\left(1+|\xi|^2\right)^2}
\end{equation}
To prove it, we make use of the recurrence relations (3.85) or (3.88). As $|M_0|^2=\left(1+|\xi|^2\right)^{N-1}$, then
\begin{equation}
	\partial_\xi\partial_{\bar{\xi}}ln|M_0|^2=(N-1)\left(1+|\xi|^2\right)^{-2}
\end{equation}
and by using relation (3.89) for $k=0$, we obtain
\begin{eqnarray}
	\partial_\xi\partial_{\bar{\xi}}ln|M_0|^2&=&|M_1|^2|M_0|^{-2}-|M_0|^2|M_{-1}|^{-2} \nonumber \\
	&=&\frac{|M_1|^2}{|M_0|^2}
\end{eqnarray}
where by definition $M_{-1}=0$. From equations (3.94) and (3.95), we have
\begin{equation}
	\frac{|M_1|^2}{|M_0|^2}=\frac{(N-1)}{\left(1+|\xi|^2\right)^2}
\end{equation}
For general $k,\ 1<k<(N-1)$, we use inductive proof by assuming
\begin{equation}
	\frac{|M_k|^2}{|M_{k-1}|^2}=\frac{k(N-k)}{\left(1+|\xi|^2\right)^2}
\end{equation}
which is already true for $k=1$. We can write $|M_k|^2$ using equation (3.97) as
\begin{eqnarray}
	|M_k|^2&=&\frac{|M_k|^2}{|M_{k-1}|^2}\frac{|M_{k-1}|^2}{|M_{k-2}|^2}\ldots\frac{|M_1|^2}{|M_0|^2}|M_0|^2 \nonumber \\
	&=&\frac{k!(N-1)!}{(N-k-1)!}\left(1+|\xi|^2\right)^{N-2k-1}
\end{eqnarray}
then if we take differential over $\xi$ and $\bar{\xi}$ of equation (3.98), we obtain
\begin{equation}
	\partial_\xi\partial_{\bar{\xi}}ln|M_0|^2=\frac{(N-2k-1)}{\left(1+|\xi|^2\right)^2}
\end{equation}
Substitute equations (3.97) and (3.99) into (3.89), gives
\begin{equation}
	\frac{(N-2k-1)}{\left(1+|\xi|^2\right)^2}=\frac{|M_{k+1}|^2}{|M_k|^2}-\frac{k(N-k)}{\left(1+|\xi|^2\right)^2}
\end{equation}
therefore
\begin{equation}
	\frac{|M_{k+1}|^2}{|M_k|^2}=\frac{(k+1)(N-k-1)}{\left(1+|\xi|^2\right)^2}
\end{equation}
is true.

\textbf{Corollary}: Each configuration $M_k,\ (k=0,\ldots,N-1)$, of the Veronese map (3.92), has
\begin{equation}
	\mathcal{N}_k=(k+1)(N-k-1)
\end{equation}
We prove it using the integral formula on surface
\begin{eqnarray}
	i\int \frac{d\xi d\bar{\xi}}{\left(1+|\xi|^2\right)^2}&=&i\int det\left|\begin{array}{cc}	e^{i\phi} & i\rho e^{i\phi} \\	e^{-i\phi} & -i\rho e^{i\phi}\end{array}\right| \frac{d\rho d\phi}{(1+\rho^2)^2}\nonumber \\
	&=&2\pi	
\end{eqnarray}
with $\rho=tan\left(\frac{\theta}{2}\right)$. Substitute equations (3.103) and (3.101) into (3.78) for $CP^{N-1}$ case, we obtain
\begin{equation}
	\mathcal{N}_k=(k+1)(N-k-1)
\end{equation}

\chapter{$SU(N)$ Bogomolny Solutions}
\section{Harmonic Maps Ansatz}
In this chapter, we discuss the general constraction of the solutions for
$SU(N)$ Bogomolny equations for BPS magnetic monopoles using previous
description in harmonic maps. We also give an example for $SU(2)$ Bogomolny
equations that give the same solutions with one was proved by M.K. Prasad and
C.M. Sommerfield~\cite{prasad}.

For this purpose, we take an ansatz for $SU(N)$ Bogomolny equations
given by T. Ioannidou and P.M. Sutcliffe~\cite{ioannidou2}
\begin{equation}
	\Phi=i\sum_{j=0}^{N-2}h_j(r)\left(P_j-\frac{1}{N}\right),\ A_\xi=\sum_{j=0}^{N-2}g_j(r)\left[P_j,\partial_\xi P_j\right],\ A_r=0
\end{equation}
We substitute the ansatz into $SU(N)$ Bogomolny equations (2.47) and (2.48) then multiply
from the right with vectors field $M_l$ as follows
\begin{eqnarray}
	(iD_\xi\Phi-F_{r\xi})M_l=0 \\
	\left(iD_r\Phi-\frac{\left(1+|\xi|^2\right)^2}{2r^2}F_{\xi\bar{\xi}}\right)M_l=0
\end{eqnarray}
Next, the equation (4.2) is expressed in term of projection operators $P_k$. In that case, we need to describe the properties of projection operators $P_k$ and their derivatives applied to $M_l$:
\begin{equation}
	P_kM_l=\delta_{kl}M_l
\end{equation}
from equations (3.59) and (3.60), we obtain
\begin{eqnarray}
	\partial_\xi P_kM_l&=&\partial_\xi\left(M_k|M_k|^{-2}\right)M_k^\dagger M_l+M_k|M_k|^{-2}\partial_\xi M_k^\dagger M_l \nonumber \\
	&=&\left(\delta_{kl}-\delta_{k,l+1}\right)M_{l+1}
\end{eqnarray}
\begin{eqnarray}
	\partial_{\bar{\xi}} P_kM_l&=&\partial_{\bar{\xi}}(P_kM_l) - P_k\partial_{\bar{\xi}}M_l \nonumber \\
	&=&\left(\delta_{k,l-1}-\delta_{kl}\right)M_{l-1}\mathcal{K}_{l-1}
\end{eqnarray}
where $\mathcal{K}_{l-1}=|M_{l-1}|^{-2}|M_l|^2$. For simplicity, we use a convention for index $j$ which means the sum of $j=0,\ldots,N-2$.

Therefore, we can take the action of the fields, as given by the ansatz (4.1), on the vectors field $M_l$ by using properties (4.4)-(4.6)
\begin{eqnarray}
	A_\xi M_l=-G_lM_{l+1} \\
	A_{\bar{\xi}}M_l=G_{l-1}M_{l-1}\mathcal{K}_{l-1}
\end{eqnarray}
where $G_l=g_l+g_{l-1}$. Then, we derive that
\begin{eqnarray}
	F_{r\xi}M_l=-\partial_rG_lM_{l+1} \\
	F_{\xi\bar{\xi}}M_l=\left(\partial_\xi A_{\bar{\xi}}-\partial_{\bar{\xi}} A_\xi+\left[A_\xi,A_{\bar{\xi}}\right]\right)M_l
\end{eqnarray}
For equation (4.10), we calculate separately for its parts as follows
\begin{eqnarray}
	\partial_\xi A_{\bar{\xi}}M_l=M_l(\mathcal{K}_{l-1}G_{l-1}-\mathcal{K}_lG_l)\\
	\partial_{\bar{\xi}}A_\xi M_l=M_l(\mathcal{K}_lG_l-\mathcal{K}_{l-1}G_{l-1})\\
	A_\xi A_{\bar{\xi}}M_l=-G_{l-1}^2M_l\mathcal{K}_{l-1}\\
	A_{\bar{\xi}}A_\xi M_l=-G_l^2M_l\mathcal{K}_l
\end{eqnarray}
Substitute equations (4.11)-(4.14) into equation (4.10), brings
\begin{equation}
F_{\xi\bar{\xi}}M_l=M_l\left(\mathcal{K}_{l-1}\widetilde{\mathcal{Q}}_{l-1}-\mathcal{K}_l\widetilde{\mathcal{Q}}_l\right)
\end{equation}
where $\widetilde{\mathcal{Q}}_l=G_l(2-G_l)$, then
\begin{eqnarray}
	\Phi M_l=i\left(h_l-\frac{h_j}{N}\right)M_l\\
	\partial_r\Phi M_l=i\left(\partial_rh_l-\frac{\partial_rh_j}{N}\right)M_l\\
	\partial_\xi\Phi M_l=i(h_l-h_{l+1})M_{l+1}
\end{eqnarray}
Substitute all the previous result into equation (4.2), gives
\begin{eqnarray}
	\{i(\partial_\xi\Phi+[A_\xi,\Phi])-F_{r\xi}\}M_l=0 \nonumber \\
	\{(h_{l+1}-h_l)-(h_{l+1}-h_l)G_l+\partial_rG_l\}M_{l+1}=0
\end{eqnarray}
because $M_{l+1}\neq 0$, therefore
\begin{equation}
	(h_{l=1}-h_l)(1-G_l)+\partial_rG_l=0
\end{equation}
If we do the same way to the equation (4.3), then we have
\begin{eqnarray}
	\left(i\partial_r\Phi-\frac{(1+|\xi|^2)^2}{2r^2}F_{\xi\bar{\xi}}\right)M_l=0 \nonumber \\
	-\left(\partial_rh_l-\frac{\partial_rh_j}{N}\right)M_l-\frac{(1+|\xi|^2)^2}{2r^2}M_l(\mathcal{K}_{l-1}\widetilde{\mathcal{Q}}_{l-1}-\mathcal{K}_l\widetilde{\mathcal{Q}}_l)=0
\end{eqnarray}
For $CP^{(N-1)}$ case, we take vector fields $M_l$ as given by the Veronese map from the previous chapter
\begin{equation}
	M_0=\left[1,\sqrt{C_1^{N-1}}\xi,\ldots,\sqrt{C_k^{N-1}}\xi^k,\ldots,\xi^{N-1}\right]^T
\end{equation}
then we may write the factors $(1+|\xi|^2)^2\mathcal{K}_l$ as constant $K_l$, such that
\begin{equation}
	\mathcal{K}_l=\frac{K_l}{(1+|\xi|^2)^2}
\end{equation}
where we find that $K_l=\mathcal{N}_l=(l+1)(N-l-1)$ are constants which depend on
index $l$ as shown from the equation (3.100). Hence, the equation (4.21) becomes
\begin{equation}
	\left\{\left(\frac{\partial_rh_j}{N}-\partial_rh_l\right)-\frac{1}{2r^2}\left(K_{l-1}\widetilde{\mathcal{Q}}_{l-1}-K_l\widetilde{\mathcal{Q}}_l\right)\right\}M_l=0
\end{equation}
because $M_l\neq 0$, then we obtain
\begin{equation}
\left(\frac{\partial_rh_j}{N}-\partial_rh_l\right)-\frac{1}{2r^2}\left(K_{l-1}\widetilde{\mathcal{Q}}_{l-1}-K_l\widetilde{\mathcal{Q}}_l\right)=0
\end{equation}
or in complete
\begin{equation}
	\left(\frac{\sum_{j=0}^{N-2}\partial_rh_j}{N}-\partial_rh_l\right)-\frac{1}{2r^2}\left(K_{l-1}\widetilde{\mathcal{Q}}_{l-1}-K_l\widetilde{\mathcal{Q}}_l\right)
\end{equation}
Note that, by definition $h_l,g_l=0$ if $l\notin\{0,1,\ldots,(N-2)\}$.

The equations (4.20) and (4.26) are called $SU(N)$ Bogomolny equations for
BPS magnetic monopoles. Clearly, those equations are in of scalar and non-linear
coupled fields. Compared to the form of the $SU(N)$ Bogomolny equations before
we apply harmonic maps method, the previous $SU(N)$ Bogomolny equations are in
form of matrix and aslo non-linear coupled fields. It is much easy to work with
scalar field equations than in form of matrix field equations. But, to find the exact
solutions of $SU(N)$ Bogomolny equations, even in scalar field equations, is
another complex problem since the fields in form of non- linear coupled. In this
thesis, we do not discuss about the exact solutions since it needs others advance
mathematical techniques to find the soliton solutions.

\section{$SU(2)$ Bogomolny Equation}
In this section, we show a simple example for $SU(2)$ Bogomolny case for which the equations (4.20) and (4.26) has range of value $l=0,\ldots,(2-2)=0$, so by definition $h_l,g_l=0$ if $l\neq 0$, then equations (4.20) and (4.26) become
\begin{eqnarray}
	-h_0(1-g_0)+\dot{g}_0=0 \\
	-\frac{1}{2}\dot{h}_0+\frac{1}{2r^2}g_0(2-g_0)=0
\end{eqnarray}
with $\dot{g}_0=\partial_rg_0$ and $\dot{h}_0=\partial_rh_0$

Now, define a function $K=1-g_0$ and substitute it into equations (4.27) and (4.28), such that
\begin{eqnarray}
	-h_0K-\dot{K}=0\\
	-\dot{h}_0+\frac{1}{r^2}(1-K^2)=0
\end{eqnarray}
From equation (4.30), we have
\begin{equation}
	\dot{h}_0=\frac{1}{r^2}(1-K^2)
\end{equation}
Take the differential of equation (4.29) about $r$ and substitute it into equation
(4.31)
\begin{eqnarray}
	-\dot{h}_0K-h_0\dot{K}-\ddot{K}=0 \nonumber \\
	\ddot{K}=-\frac{1}{r^2}K(1-K^2)-h_0\dot{K}
\end{eqnarray}
from equation (4.29), gives
\begin{equation}
	r^2\ddot{K}=K(K^2-1)+r^2h_0^2K
\end{equation}
Take the differential of equation (4.30) about $r$ and substitute it into
equation (4.29), then
\begin{eqnarray}
	-\ddot{h}_0+\frac{1}{r^2}(-2K\dot{K})-\frac{2}{r^3}(1-K)=0 \nonumber \\
	r^2\ddot{h}=-\frac{2}{r}(1-K)+2h_0K^2
\end{eqnarray}
Next, we define a function $H=rh_0$ with
\begin{eqnarray}
	\dot{h}_0=-\frac{H}{r^2}+\frac{\dot{H}}{r} \\
	\ddot{h}_0=\frac{2H}{r^3}-\frac{2\dot{H}}{r^2}+\frac{\ddot{H}}{r}
\end{eqnarray}
and substitute into equations (4.33) and (4.34)
\begin{eqnarray}
	r^2\ddot{K}=K(K^2-1)+H^2K \\
	r^2\ddot{H}=2K^2H
\end{eqnarray}
The result in equations (4.37) and (4.38) are same as the one that was obtained by
M. Prasad and C. Sommerfield~\cite{prasad} which gives
\begin{eqnarray}
	K=\frac{C\ r}{\sinh{(C\ r)}} \\
	H=C\ r\coth{(C\ r)}-1
\end{eqnarray}
where $C$ is a constant.

\chapter{Conclusions}
In this thesis, we have described how to use harmonic maps in $CP^{(N-1)}$
space to simplify the $SU(N)$ Bogomolny equations for BPS magnetic monopoles.
As the result obtained in chapter IV equations (4.20) and (4.26), we just need to
solve the non-linear coupled equations for scalar fields that depend on variable $r$,
as follows:
\begin{eqnarray}
	(h_{l+1}-h_l)(1-G_l)+\partial_rG_l=0 \\
\left(\frac{\sum^{N-2}_{j=0}\partial_rh_j}{N}-\partial_rh_l\right)-\frac{1}{2r^2}(K_{l-1}\widetilde{\mathcal{Q}}_{l-1}-K_l\widetilde{\mathcal{Q}}_l)=0
\end{eqnarray}
with $l=0,1,\ldots,(N-2)$.

Even it is not a simple problem to solve the non-linear coupled equations, but at
least we have more simple form of the equations in scalar than before. As an
example, we also have shown for $SU(2)$ Bogomolny equations and verified the
result with the one obtained by M. Prasad and C. Sommerfield.

%****************************************** A P E N D I C E   A **************************************************

\appendix
\chapter{Riemann Sphere}
In deriving metric (2.19), we introduce equation $\xi=\tan{(\frac{\theta}{2})}e^{i\phi}$ and transform variables of the metric (2.18) into complex variables $(\theta,\phi)\rightarrow(\xi,\bar{\xi})$. By using other forms of the
transformation equation which are
\begin{eqnarray}
	\tan{\left(\frac{\theta}{2}\right)}=\sqrt{\xi\bar{\xi}}=|\xi|,\ \sin{\left(\frac{\theta}{2}\right)}=\frac{|\xi|}{\sqrt{1+|\xi|^2}},\ \cos{\left(\frac{\theta}{2}\right)}=\frac{1}{\sqrt{1+|\xi|^2}}
\end{eqnarray}
,so that
\begin{eqnarray}
	\partial_\xi\theta=\frac{|\xi|}{(1+|\xi|^2)\xi},\ \partial_{\bar{\xi}}\theta=\frac{|\xi|}{(1+|\xi|^2)\bar{\xi}} \nonumber \\
	\partial_\xi\phi=-\frac{i}{2\xi},\ \partial_{\bar{\xi}}\phi=\frac{i}{2\bar{\xi}}
\end{eqnarray}
with the result of equations (A.1) and (A.2), then $d\theta^2$ and $d\phi^2$ are written as
\begin{eqnarray}
	\left(\partial_\xi\theta\ d\xi+\partial_{\bar{\xi}}\theta\ d\bar{\xi}\right)^2=\frac{|\xi|^2}{(1+|\xi|^2)^2}\left(\frac{1}{\xi^2}d\xi^2+\frac{1}{\bar{\xi}^2}d\bar{\xi}^2\right)+\frac{1}{(1+|\xi|^2)^2}(d\xi d\bar{\xi}+d\bar{\xi}d\xi) \nonumber \\
	\left(\partial_\xi\phi\ d\xi+\partial_{\bar{\xi}}\phi\ d\bar{\xi}\right)^2=-\frac{1}{4}\left(\frac{1}{\xi^2}d\xi^2+\frac{1}{\bar{\xi}^2}d\bar{\xi}^2-\frac{1}{|\xi|^2}(d\xi d\bar{\xi}+d\bar{\xi}d\xi)\right)
\end{eqnarray}
and
\begin{eqnarray}
	\sin^2\theta&=&4\sin^2\left(\frac{\theta}{2}\right)\cos^2\left(\frac{\theta}{2}\right) \nonumber \\
	&=&\frac{4|\xi|^2}{(1+|\xi|^2)^2}
\end{eqnarray}
Substitute all the result from (A.3) and (A.4) into equations (2.18) then we have metric (2.19)
\begin{equation}
	ds^2=dt^2-dr^2-\frac{2r^2}{(1+|\xi|^2)^2}(d\xi d\bar{\xi}+d\bar{\xi}d\xi)
\end{equation}
So, we can write its tensor metric by
\begin{eqnarray}
	g_{\mu\nu}=\left(\begin{array}{cccc}
	1 & 0 & 0 & 0 \\
	0 & -1 & 0 & 0 \\
	0 & 0 & 0 & -\frac{2r^2}{(1+|\xi|^2)^2} \\
	0 & 0 & -\frac{2r^2}{(1+|\xi|^2)^2} & 0
	\end{array}\right) \\
	g^{\mu\nu}=\left(\begin{array}{cccc}
	1 & 0 & 0 & 0 \\
	0 & -1 & 0 & 0 \\
	0 & 0 & 0 & -\frac{(1+|\xi|^2)^2}{2r^2} \\
	0 & 0 & -\frac{(1+|\xi|^2)^2}{2r^2} & 0
	\end{array}\right)
\end{eqnarray}

\chapter{Derivation of Equations (3.40) and (3.41)}
To derive equation (3.40), we need to use tensor metric (A.7), such that equation (3.39) becomes
\begin{equation}
	[P,\partial^\mu\partial_\mu P]=[P,g^{\mu\nu}\partial_\mu\partial_\nu P]
\end{equation}
As stated in section 3..4.2 that the source manifold is a two dimensional Euclidean
space which is wrote is Riemann sphere coordinates. It means that tensor metric
(A.7) only use variables of Riemann sphere $(\xi,\bar{\xi})$, then the equation(B.1) becomes
\begin{eqnarray}
	[P,\partial^\mu\partial_\mu P]&=&[P,g^{\xi\bar{\xi}}\partial_\xi\partial_{\bar{\xi}}P+g^{\bar{\xi}\xi}\partial_{\bar{\xi}}\partial_\xi P]\nonumber \\
	&=&[P,-\frac{(1+|\xi|^2)^2}{r^2}\partial_\xi\partial_{\bar{\xi}}P] \nonumber \\
	&=&-\frac{(1+|\xi|^2)^2}{r^2}[P,\partial_\xi\partial_{\bar{\xi}}P] 	
\end{eqnarray}
and from equation (3.39) then
\begin{equation}
	[P,\partial_\xi\partial_{\bar{\xi}}P]=0
\end{equation}
Next, for equation (3.41), we write equation (3.40) as follows:
\begin{eqnarray}
	[P,\partial_\xi\partial_{\bar{\xi}}P]+[P,\partial_\xi\partial_{\bar{\xi}}P]=0 \nonumber \\
	\partial_\xi(P\partial_{\bar{\xi}}P)-\partial_\xi(\partial_{\bar{\xi}}PP)+\partial_{\bar{\xi}}(P\partial_\xi P)-\partial_{\bar{\xi}}(\partial_\xi PP)=0 \nonumber \\
	\partial_\xi[P,\partial_{\bar{\xi}}P]+\partial_{\bar{\xi}}[P,\partial_\xi P]=0
\end{eqnarray}

\chapter{Derivation of Equations (3.49) and (3.50)}
Equation (3.49) is derived by writing it in form of matrix field $M$:
\begin{eqnarray}
	(I-P)M&=&M-M|M|^{-2}M^\dagger M \nonumber \\
	&=&0
\end{eqnarray}
and if we take its Hermitian conjugate, then
\begin{eqnarray}
	((I-P)M)^\dagger&=&0 \nonumber \\
	M^\dagger(I-P)&=&0	
\end{eqnarray}
While for equation (3.49), we can derive it from instanton solution as
follows:
\begin{eqnarray}
	P\partial_\xi P&=&0 \nonumber \\
	P\partial_\xi (I-P)&=&0 \nonumber \\
	\partial_\xi(P (I-P))-\partial_\xi P\ (I-P)&=&0 \nonumber \\
	\partial_\xi(M|M|^{-2}M^\dagger (I-P))-\partial_\xi(M|M|^{-2}M^\dagger)(I-P)&=&0 \nonumber \\
	-\partial_\xi(M|M|^{-2})M^\dagger (I-P)-M|M|^{-2}\partial_\xi M^\dagger(I-P)&=&0 \nonumber \\
	M|M|^{-2}\partial_\xi M^\dagger(I-P)&=&0 \nonumber \\
	M|M|^{-2}(\partial_{\bar{\xi}}M)^\dagger(I-P)&=&0
\end{eqnarray}

\chapter{Derivation of Properties (3.59) and (3.60)}
In this Appendix, we present derivation of the properties of equations
(3.59) and (3.60). For equation (3.59), we use the fact that the sequence of $M_k$ are independent, so we have the expansion:
\begin{equation}
	\partial_{\bar{\xi}}M_l=\sum_{k=0}^{\lambda}M_ka_{kl},\ l=1,\ldots,\lambda
\end{equation}
where $A_{kl}$ are the $(n\times n)$ matrices. We have an assumption that $M_0$ is holomorphic $\partial_{\bar{\xi}}M_0=0$.

Multiplying (D.1) from the left by $M_m^\dagger$ then we
have
\begin{equation}
	M_m^\dagger\partial_{\bar{\xi}}M_l=|M_m|^2a_{ml}
\end{equation}
From recurrence relation (3.56), gives
\begin{eqnarray}
	M_{m+1}&=&(I-P_m)\partial_xi M_m \nonumber \\
	M_{m+1}^\dagger &=&(\partial_\xi M_m)^\dagger (I-P_m) \nonumber \\
	(\partial_\xi M_m)^\dagger &=& M_{m+1}^\dagger+(\partial_\xi M_m)^\dagger P_m
\end{eqnarray}
and the left side of equation (D.2) becomes
\begin{eqnarray}
	M_m^\dagger\partial_{\bar{\xi}}M_l&=&\partial_{\bar{\xi}}(M_m^\dagger M_l)-\partial_{\bar{\xi}}M_m^\dagger M_l \nonumber \\
	&=&\partial_{\bar{\xi}}(M_l^\dagger M_l\delta_{lm})-(\partial_{\xi}M_m)^\dagger M_l \nonumber \\
	&=&\delta_{lm}M_l^\dagger\partial_{\bar{\xi}}M_l+\delta_{lm}\partial_{\bar{\xi}}M_l^\dagger M_l-(\partial_{\xi}M_m)^\dagger M_l
\end{eqnarray}
Substitute equation (D.3) into (D.4), such that
\begin{eqnarray}
M_m^\dagger\partial_{\bar{\xi}}M_l&=&\delta_{lm}M_l^\dagger\partial_{\bar{\xi}}M_l+\delta_{lm}\partial_{\bar{\xi}}M_l^\dagger M_l-(M_{l+1}^\dagger+(\partial_{\xi}M_m)^\dagger P_m)^\dagger M_l \nonumber \\
&=&\delta_{lm}M_l^\dagger\partial_{\bar{\xi}}M_l+\delta_{lm}\partial_{\bar{\xi}}M_l^\dagger M_l-\delta_{l,m+1}|M_l|^2-\partial_{\bar{\xi}}M_m^\dagger M_l\delta_{lm} \nonumber \\
&=&\delta_{lm}M_l^\dagger\partial_{\bar{\xi}}M_l+\delta_{lm}\partial_{\bar{\xi}}M_l^\dagger M_l-\delta_{l,m+1}|M_l|^2-\delta_{lm}\partial_{\bar{\xi}}M_l^\dagger M_l \nonumber \\
&=&\delta_{lm}M_l^\dagger\partial_{\bar{\xi}}M_l-\delta_{l,m+1}|M_l|^2
\end{eqnarray}
and substitute the result into equation (D.2) as
\begin{eqnarray}
	M_m^\dagger\partial_{\bar{\xi}}M_l&=&|M_m|^2a_{ml} \nonumber \\
	&=&\delta_{lm}M_l^\dagger\partial_{\bar{\xi}}M_l-\delta_{l,m+1}|M_l|^2 \nonumber \\
	a_{ml}&=&\delta_{lm}|M_m|^{-2}M_l^\dagger\partial_{\bar{\xi}}M_l-\delta_{l,m+1}|M_m|^{-2}|M_l|^2
\end{eqnarray}
If we substitute equation (D.6) into (D.1), we obtain
\begin{eqnarray}
	\partial_{\bar{\xi}}M_l&=&\sum_{k=0}^{\lambda}M_ka_{kl} \nonumber \\
&=&\sum_{k=0}^{\lambda}M_k\left(\delta_{lk}|M_k|^{-2}M_l^\dagger\partial_{\bar{\xi}}M_l-\delta_{l,k+1}|M_k|^{-2}|M_l|^2\right) \nonumber \\
&=&M_l|M_l|^{-2}M_l^\dagger\partial_{\bar{\xi}}M_l-M_{l-1}|M_{l-1}|^{-2}|M_l|^2
\end{eqnarray}
Because $a_{ll}=|M_l|^{-2}M_l^\dagger\partial_{\bar{\xi}}M_l$ and $a_{l-1,l}=-|M_{l-1}|^{-2}|M_l|^2$, then
\begin{equation}
	\partial_{\bar{\xi}}M_l=M_la_{ll}+M_{l-1}a_{l-1,l}
\end{equation}
Next, we prove that $a_{ll}=0,\ l=1,\ldots,\lambda$ or equivalent to $M_l^\dagger\partial_{\bar{\xi}}M_l=0$. For the case $l=1$, using the construction (3.57) or (3.58), we obtain
\begin{eqnarray}
	M_1&=&(I-P_0)\partial_\xi M_0,\ \partial_{\bar{\xi}}M_0=0 \nonumber \\
	\partial_{\bar{\xi}}M_l&=&\partial_{\bar{\xi}}(I-P_0)\partial_\xi M_0+(I-P_0)\partial_\xi\partial_{\bar{\xi}}M_0 \nonumber \\
	&=&-\partial_{\bar{\xi}}P_0\partial_\xi M_0 \nonumber \\
	&=&-\partial_{\bar{\xi}}(M_0|M_0|^{-2}M_0^\dagger)\partial_\xi M_0 \nonumber \\
	&=&-M_0\partial_{\bar{\xi}}(|M_0|^{-2}M_0^\dagger)\partial_\xi M_0 \nonumber \\
	M_l^\dagger\partial_{\bar{\xi}}M_l&=&0
\end{eqnarray}
For general case with $1<k<\lambda$, we use the inductive argument by assuming that $a_{kk}=0$ such that $\partial_{\bar{\xi}}M_k=M_{k-1}a_{k-1,k}$.

As
\begin{eqnarray}
	\partial_{\bar{\xi}}P_k&=&\partial_{\bar{\xi}}M_k|M_k|^{-2}M_k^\dagger+M_k\partial_{\bar{\xi}}(|M_k|^{-2}M_k^\dagger) \nonumber \\
	&=&(\partial_{\bar{\xi}}M_k+M_k\partial_{\bar{\xi}})(|M_k|^{-2}M_k^\dagger) \nonumber \\
	&=&(M_{k-1}a_{k-1,k}+M_k\partial_{\bar{\xi}})(|M_k|^{-2}M_k^\dagger)
\end{eqnarray}
using the construction (3.59) and orthogonality property (3.51), it follows that
\begin{eqnarray}
	M_{k+1}&=&(I-P_0-\cdots-P_k)\partial_\xi^{k+1}M_0 \nonumber \\
	\partial_{\bar{\xi}}M_{k+1}&=&\partial_{\bar{\xi}}(I-P_0-\cdots-P_k)\partial_\xi^{k+1}M_0+(I-P_0-\cdots-P_k)\partial_\xi^{k+1}\partial_{\bar{\xi}}M_0 \nonumber \\
	&=&-\partial_{\bar{\xi}}(P_0+\cdots+P_k)\partial_\xi^{k+1}M_0 \nonumber \\
	&=&-\sum_{l=0}^{k}\partial_{\bar{\xi}}P_l\partial_\xi^{k+1}M_0
\end{eqnarray}
If we multiply equation (D.11) from the left by $M_{k+1}^\dagger$
\begin{eqnarray}
	M_{k+1}^\dagger M_{k+1}&=&-M_{k+1}^\dagger\sum_{l=0}^{k}\partial_{\bar{\xi}}P_l\partial_\xi^{k+1}M_0 \nonumber \\
&=&-\sum_{l=0}^{k}M_{k+1}^\dagger(M_{l-1}a_{l-1,l}+M_l\partial_{\bar{\xi}})(|M_l|^{-2}M_l^\dagger)\partial_\xi^{k+1}M_0 \nonumber \\
	&=&0
\end{eqnarray}
then it shows that $a_{k+1,k+1}=0$. Therefore, equation (D.8) becomes
\begin{eqnarray}
	\partial_{\bar{\xi}}M_l&=&M_la_{ll}+M_{l-1}a_{l-1,l} \nonumber \\
	&=&M_{l-1}a_{l-1,l} \nonumber \\
	&=&-M_{l-1}|M_{l-1}|^{-2}|M_l|^2
\end{eqnarray}
which is proof of equation (3.59).

We may write equation (D.13) as follows:
\begin{equation}
	\partial_\xi M_k^\dagger=(a_{k-1,k})^\dagger M_{k-1}^\dagger
\end{equation}
and use the orthogonality property (3.50), then
\begin{eqnarray}
	\partial_\xi|M_k|^2&=&M_k^\dagger\partial_\xi M_k+\partial_\xi M_k^\dagger M_k \nonumber \\
	&=&M_k^\dagger\partial_\xi M_k+(a_{k-1,k})^\dagger M_{k-1}^\dagger M_k \nonumber \\
	&=&M_k^\dagger\partial_\xi M_k
\end{eqnarray}
such that
\begin{equation}
	\partial_\xi(M_k|M_k|^{-2})=\partial_\xi M_k|M_k|^{-2}+M_k\partial_\xi|M_k|^{-2}
\end{equation}
We have
\begin{eqnarray}
	\partial_\xi(|M_k|^2|M_k|^{-2})&=&0 \nonumber \\
	|M_k|^2\partial_\xi|M_k|^{-2}+\partial_\xi|M_k|^2|M_k|^{-2}&=&0 \nonumber \\
	\partial_\xi |M_k|^{-2}&=&-|M_k|^{-2}\partial_\xi|M_k|^2|M_k|^{-2}
\end{eqnarray}
and substitute it into equation (D.16), so that
\begin{eqnarray}
	\partial_\xi(M_k|M_k|^{-2})&=&\partial_\xi M_k|M_k|^{-2}-M_k|M_k|^{-2}\partial_\xi|M_k|^2|M_k|^{-2} \nonumber \\
	&=&\partial_\xi M_k|M_k|^{-2}-M_k|M_k|^{-2}M_k^\dagger\partial_\xi M_k |M_k|^{-2} \nonumber \\
	&=&(I-P_k)\partial_\xi M_k|M_k|^{-2} \nonumber \\
	&=&M_{k+1}|M_k|^{-2}
\end{eqnarray}
which is proof of equation (3.60).

\chapter{Derivation of Equations (3.82) and (3.84)}
Derivation of equation (3.82) works as follows:
\begin{eqnarray}
	|M_k|^\dagger\partial_\xi\partial_{\bar{\xi}}M_k|M_k|^{-2}&=&M_k^\dagger\partial_\xi\left(-M_{k-1}|M_{k-1}|^{-2}|M_k|^2\right)|M_k|^{-2} \nonumber \\
	&=&-\partial_\xi\left(M_k^\dagger M_{k-1}|M_{k-1}|^{-2}|M_k|^2\right)|M_k|^{-2} \nonumber \\
	&&+(\partial_{\bar{\xi}}M_k)^\dagger M_{k-1}|M_{k-1}|^{-2}|M_k|^2|M_k|^{-2} \nonumber \\
	&=&-|M_k|^2|M_{k-1}|^{-2}M_{k-1}^\dagger M_{k-1}|M_{k-1}|^{-2} \nonumber \\
	&=&-|M_k|^2|M_{k-1}|^{-2}
\end{eqnarray}
While for equation (3.84), we derive as follows:
\begin{eqnarray}
	M_k^\dagger\partial_{\bar{\xi}}(P_k\partial_\xi M_k)|M_k|^{-2}&=&\partial_{\bar{\xi}}(M_k^\dagger P_k\partial_\xi M_k)|M_k|^{-2}-\partial_{\bar{\xi}}M_k^\dagger P_k\partial_\xi M_k|M_k|^{-2} \nonumber \\
	&=&\partial_{\bar{\xi}}(M_k^\dagger\partial_\xi M_k)|M_k|^{-2}-\partial_{\bar{\xi}}M_k^\dagger M_k|M_k|^{-2}M_k^\dagger\partial_\xi M_k|M_k|^{-2} \nonumber \\
	&=&\partial_{\bar{\xi}}\left(\partial_\xi(M_k^\dagger M_k)-\partial_\xi M_k^\dagger M_k\right)|M_k|^{-2} \nonumber \\
	&=&\partial_\xi\partial_{\bar{\xi}}(M_k^\dagger M_k)|M_k|^{-2} \nonumber \\
	&=&\partial_\xi\left(\partial_{\bar{\xi}}|M_k|^2|M_k|^{-2}\right)-\partial_{\bar{\xi}}(M_k^\dagger M_k)\partial_\xi|M_k|^{-2} \nonumber \\
	&=&\partial_\xi\left(\partial_{\bar{\xi}}|M_k|^2|M_k|^{-2}\right)-\partial_{\bar{\xi}}M_k^\dagger M_k\partial_\xi|M_k|^{-2}-M_k^\dagger\partial_{\bar{\xi}}M_k\partial_\xi|M_k|^{-2} \nonumber \\
	&=&\partial_\xi\left(\partial_{\bar{\xi}}|M_k|^2|M_k|^{-2}\right)
\end{eqnarray}

\backmatter

%****************************************** A G R A D E C I M I E N T O S *********************************************

%\input{agradecimiento}

%****************************************** R E F E R E N C I A S **************************************************
\bibliographystyle{unsrt}

%%%%\bibliographystyle{unsrt}%%%

%%%\bibliography{thesis}%%%%

\end{document}